\documentclass[preprint,12pt]{elsarticle}

\usepackage[utf8]{inputenc}
\usepackage[english]{babel}
\usepackage{graphics,graphicx}
\usepackage{float}
\usepackage{geometry}
\usepackage{setspace}
\usepackage[usenames,dvipsnames]{color}
\usepackage[dvipsnames]{xcolor}

\usepackage[fleqn]{amsmath,mathtools}
\usepackage{amsxtra,amsfonts,latexsym,amssymb,euscript}
\usepackage{amstext}
\usepackage{mathrsfs}
\usepackage{algorithm}
\usepackage{algpseudocode}

\journal{International Journal of Solids and Structures}

\begin{document}
\begin{frontmatter}
\title{Failure through crack propagation in components with holes and notches: an experimental assessment of the phase field model}
\author[1]{R. Cavuoto}
\author[2]{P. Lenarda}
\author[3]{D. Misseroni}
\author[2]{M. Paggi}
\author[1]{D. Bigoni\corref{cor}}

\address[1]{Instabilities Lab, University of Trento, Via Mesiano 77, 38123 Italy}
\address[2]{IMT School for Advanced Studies Lucca, Piazza San Francesco 19, 55100 Lucca, Italy}
\address[3]{Laboratory for the 
Design of Reconfigurable Metamaterials $\&$ Structures, University of Trento, Via Mesiano 77, 38123 Italy \\
\vspace{1cm} {\large Dedicated to Professor Stelios  Kyriakides} }
\cortext[cor]{Corresponding author. Tel: +390461282507,  e-mail: bigoni@ing.unitn.it}

\begin{abstract}

Fracture growth in a material is strongly influenced by the presence of inhomogeneities, which deviate crack trajectories from rectilinearity and deeply affect failure. Increasing crack tortuosity is connected to enhancement of fracture toughness, while often a crack may even be stopped when it impinges  a void, which releases the stress concentration. Therefore, the determination of crack trajectories is important in the design against failure of materials and mechanical pieces. 
The recently developed phase-field approach (AT1 and AT2 models), based on a variational approach to damage localization, is believed to be particularly suited to describe complex crack trajectories. This belief is examined through a comparison between simulations and photoelastic experiments on PMMA plates, which have been designed in a new way, to highlight the effects of notches and circular holes on fracture propagation. The latter is shown to initiate  from a notch and to be strongly attracted by voids. When a void is hit, fracture is arrested, unless the void contains a notch on its internal surface, from which a new crack nucleates and propagates.
Different mechanical models are tested where fracture initiates and grows (i.) under Mode I compact tension, (ii.) four-point bending and (iii.) a tensile stress  indirectly generated during compression of samples containing a circular hole. The experiments show that the fracture propagation may be \lq designed' to develop in different tortuous paths, involving multiple arrests and secondary nucleation. 
Simulations performed with an {\it ad hoc} implemented version of the AT1 and AT2 phase-field 
methods (equipped with spectral decomposition, in which a crack is simulated as a highly localized zone of damage accumulation) are shown to be in close agreement with experiments and therefore confirm the validity of the approach and its potentialities for mechanical design.
\end{abstract}

\end{frontmatter}

\section{Introduction}
\label{Introduction}

Originated from mathematical techniques based on $\Gamma-$convergence \cite{Maso1993,Braides1998,Braides2002, Delpiero} and tailored for the approximation of free discontinuity problems \cite{Ambrosio1990,Ambrosio1992}, the phase field regularization of brittle fracture proposed by Francfort and Marigo \cite{Francfort1998} has attracted a remarkable attention within  the computational fracture mechanics community over the last decade. As compared to other computational methods for  damage and fracture simulation in materials and components, such as for instance the Crack Band Model \cite{bazant1982}, the Smeared Crack Model \cite{bazant1988}, nonlocal and diffuse damage models \cite{bazant1990,bazant1994,pijaudier}, or gradient damage models \cite{geers}, the phase field approach offers an elegant solution for problems involving linear elastic fracture mechanics. This solution is pursued through an energy minimization which, in the $\Gamma-$convergence limit, consistently reproduces the Griffith theory of fracture. The phase field approach to fracture, further analyzed in \cite{Bourdin2000,Bourdin2008}, has been applied in a considerable series of works proposing comparisons with other nonlocal damage models and discussing several detailed aspects regarding the finite element implementation  \cite{aranson2000,kuhn2008,hakim2009,Amor2009,kuhn2010,nguyen2,Ambati2014,Msekh2015,Jod2020, Wu}. In this context, it is worth recalling the fundamental contribution by Miehe and coworkers \cite{Miehe2010,Miehe2010b} that were the first to propose a robust finite element implementation of the phase field for brittle fracture, specialized to account for damage irreversibility and based on a suitable degradation mechanism able to simulate situations involving tensile stress states. 

The state-of-the-art literature on phase field models clearly shows that the approach is mature for technical applications. In this regards, Tann\'e et al. \cite{Tanne} have recently assessed the capabilities of the phase field approach (see in particular the AT1 and the AT2 models) to predict crack nucleation from V-notches and from points with stress concentrations. The results are supporting the technical applicability of the method, but all the benchmark problems were limited to pure tensile stress states, so that only the crack nucleation stage was examined. A long series of works \cite{PR2017,teresa,quintanas,teresa2,dean,kumar2021a,kumar2021b}, where the quantitative application of the methodology has been assessed for composite materials with material heterogeneities (fibers), pinpoints that the phase field method is qualitatively able to predict realistic tortuous crack patterns observed in those materials and arising from a severe rotation of the principal axes of stress. However, a systematic comparison of the phase field model predictions with experimental results under different mechanical conditions and including not only the shape of the crack trajectory, but also the force-displacement diagram, is, to the best knowledge of the present authors, still lacking. 

The present article is therefore aimed at testing the potential of the phase field approach to brittle fracture against benchmark experimental tests where the crack trajectory is complicated by the presence of notches and circular holes, which involve stress states during crack propagation evidencing a strong rotation of the principal stress axes. Crack deviations from a rectilinear path are also very important in engineering, since they usually enhance toughness of a material, so that a tortuous crack trajectory is considered the signature of a superior fracture resistance \cite{noselli,xu,mirkhalaf}. 
The deviation from rectilinearity may be produced in different situations. Curved fracture geometries and crack branching have been analyzed from a number of perspectives  
 \cite{blittencourt,hori,sumi,sumi2, Tvergaard2008}.
In particular, one of the factors producing deviations from rectilinearity of crack growth is the presence of a perturbing element, such as a defect, or a void, or an inclusion, in an otherwise uniform material. This element changes the stress state, so that asymptotic analyses \cite{valentini,movchan,movchan2,movchan3} and numerical simulations  \cite{blittencourt} 
show that the modified stress state strongly affects the trajectory of a fracture. In particular, it has been shown that soft inclusions or voids \lq attract', while stiff inclusions \lq repel', a fracture which would propagate rectilinearly in a homogeneous material. 
The attractive nature of a void on 
crack growth has also been experimentally confirmed  \cite{faber,misseroni,misseroni2} and it is known that a fracture 
intercepting a pore or a hole with smooth boundary can even be stopped, as 
the void may release the stress concentration which drives the crack growth. 

Simulations herein performed are based on our implementation of the AT1 and AT2 phase field models which, in addition to the formulation in \cite{Tanne}, accounts also for the spectral decomposition according to \cite{Miehe2010} in order to properly simulate crack growth under tensile/compressive stress states. Numerical predictions are compared with results from experiments, involving a newly designed set-up, applied to photoelastic PMMA plates, containing notches and circular holes, where the crack trajectory strongly deviates from rectilinearity and may or may not terminate against a circular hole.
In particular, Mode I compact tension and four-point bending tests are performed on notched specimens with one or two circular holes. 
In another type of test, a sample containing a circular void is compressed, so that a crack is nucleated and slowly propagates, in a direction parallel to the compression and orthogonal to the tensile stress developing at the apex of the void.

\begin{figure}[H]
 \centering
 \includegraphics[width=1\linewidth]{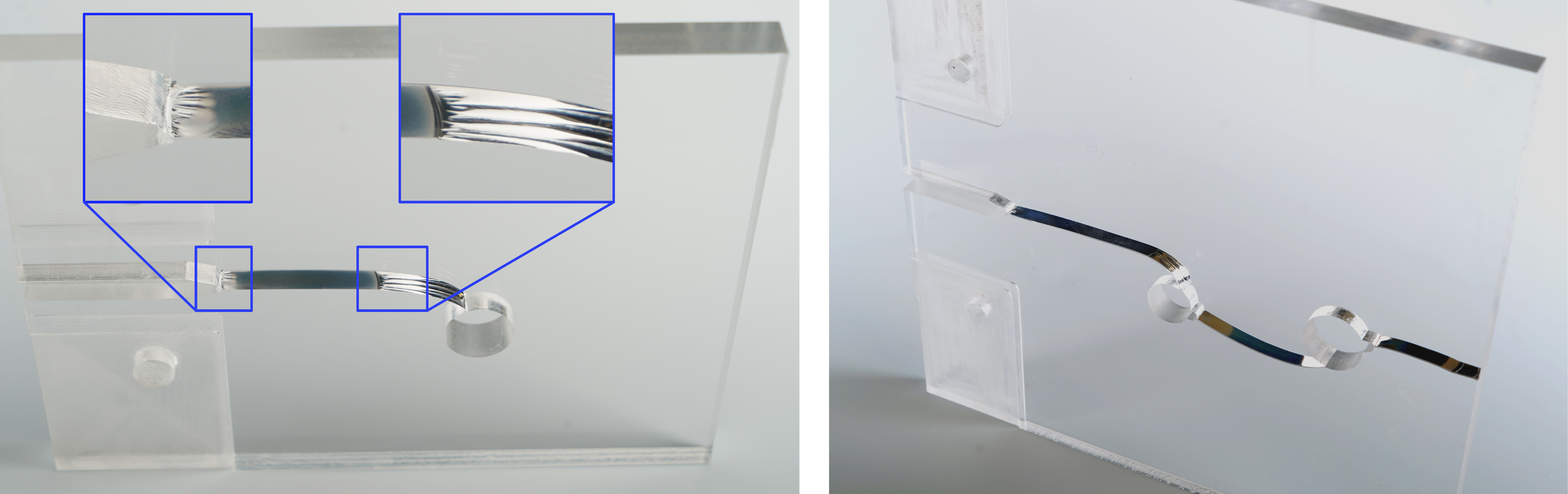}
 \caption{Post-mortem samples showing intentionally designed curved crack propagation. The crack nucleated  from a V-shaped notch and in the case on the left terminated against a circular hole, while in the case on the right a secondary crack propagated from a second notch internal to the hole, reached a second circular hole, where a third crack finally brought to failure the sample. All these behaviours have been intentionally obtained after careful design of the experiments. 
 Deviation from the initial straight path corresponds to the formation of  Wallner lines, followed by twist Hackle marks, on the crack surface (details highlighted on the left).
 }
 \label{image_intro}
\end{figure}

The new design of photoelastic experiments allows to \lq artificially' induce complex crack paths, in a way that fracture propagation may be engineered. 
For instance, 
the cracks visible in the photos reported in Fig. \ref{image_intro}, 
have been intentionally designed to hit the circular voids and to behave in a desired way.
In the case shown on the left, a fracture is deviated  from rectilinearity (this deviation corresponds to the formation of  Wallner lines, followed by twist Hackle marks, on the crack surface) before hitting a circular void, where the fracture is arrested. 
In the case shown on the right, a fracture is initially generated at a notch, is induced to impinge a first circular hole, where it is temporarily stopped. Then a second notch has been designed inside the void to stimulate a restart of propagation. This propagation has been intentionally driven toward a second void, where another designed notch re-starts crack growth, which terminates against a free boundary.  
Our experiments\footnote{
Records of photoelastic experiments performed at the Instabilities Lab of the University of Trento and documenting crack nucleation and growth, with trajectories  determined by the presence of  voids are provided as supplemental material.
} clearly 
show that the crack nucleation and growth can be intentionally conditioned to obtain, at least to a certain extent,  desired propagation features. 

This article confirms the validity of the phase-field technique to describe complex crack trajectories,  and provide quantitative predictions of structural responses in close agreement with experiments. This conclusion provides a firm experimental basis to further numerical studies dedicated to the simulation of crack propagation via phase-field in brittle materials.

\section{AT1 and AT2 phase-field approaches to fracture with spectral decomposition}
\label{Phase}

\noindent With reference to an arbitrary body occupying a domain $\Omega \in\mathbb{R}^{n_{dim}}$, with boundary $\partial\Omega\in\mathbb{R}^{n_{dim}-1}$, in the
Euclidean space of dimension $n_{dim}$, in which an evolving internal discontinuity $\Gamma$ is postulated to exist, a material point is denoted by $\mathbf{x}$ and body forces  
by $\mathbf{b}: \Omega \rightarrow\mathbb{R}^{n_{dim}}$.  Mixed  conditions are prescribed along non-overlapping regions of the boundary  $\partial\Omega_{t}\cup\partial\Omega_{u}=\partial\Omega$ in the usual form
\begin{equation}
\label{BVP}
\mathbf{u} = \overline{ \mathbf{u} } \hspace{0.2cm} \text{ on } \partial \Omega_{u}  \hspace{0.2cm} \text{ and } \hspace{0.2cm}     \boldsymbol \sigma \cdot \mathbf{n} = \mathbf{\overline{T}}  \hspace{0.2cm} \text{ on } \partial \Omega_{t} \ ,
\end{equation}
where $\mathbf{n}$ denotes the outward unit normal to the boundary, $\mathbf{u}$ is the displacement field and 
$\boldsymbol \sigma$ is the Cauchy stress tensor, while $\overline{ \mathbf{u} }$ and $\mathbf{\overline{T}}$ are prescribed surface 
displacements and tractions. 

The variational approach to brittle fracture, governing crack nucleation, propagation and branching, is set up through the definition of the free energy functional \cite{Miehe2010, Borden2014}:
\begin{equation}
\Pi (\mathbf{u}, \Gamma) =  \Pi_{\Omega} (\mathbf{u}, \Gamma) + \Pi_{\Gamma} (\Gamma) ,
\label{functional}
\end{equation}
embodying an additive decomposition between the elastic bulk energy $\Pi_{\Omega}$ stored in the damaged body and the energy $\Pi_{\Gamma}$ 
necessary to nucleate and propagate a Griffith crack \cite{Griffith1921},
defined as 
\begin{equation}
\Pi_{\Omega} (\mathbf{u}, \Gamma) + \Pi_{\Gamma} (\Gamma) =   \int_{\Omega \backslash \Gamma } \psi^{e}(\boldsymbol \varepsilon)  \ \mathrm{d}  \mathbf{x} + \int_\Gamma \mathcal{G}_c(\mathbf{u},s)\mathrm{d}\Gamma\ ,
\label{notregular}
\end{equation}
where $\psi^{e}$ is the elastic strain energy density, function of strain $\boldsymbol \varepsilon$, and $\mathcal{G}_c$ is the fracture energy, function of the displacement $\mathbf{u}$ and of the phase field variable $s$. 
The latter parameter $s \in [0,1]$ is an internal state variable, ranging between 0 and 1 and representing isotropic damage, so that $s=0$ is representative of the intact material, while $s=1$ characterizes the fully damaged state.

\subsection{The regularized variational formulation}
\label{PFbulk}

\noindent Within the regularized framework of the phase field approach \cite{Bourdin2008,Miehe2010, Delpiero, Borden2014, Arroyo, Amiri2016}, the potential energy of the system is decomposed into two terms:
\begin{equation}\label{regular}
\Pi (\mathbf{u}, s) = \int_{\Omega} \psi^e(\boldsymbol \varepsilon,s) \ \mathrm{d}  \mathbf{x}+G_c\int_\Omega
 \gamma (s, \nabla s) \ \mathrm{d}  \mathbf{x} \ ,
\end{equation}
where $\psi^{e}(\boldsymbol \varepsilon,s)$ is the energy density of the bulk, now function of the damaged parameter $s$, and  $\gamma(s, \nabla s)$ is the crack density functional, with $\nabla $ denoting the spatial gradient operator. As a result, the total free energy density of the bulk $\hat{\psi}$ reads as
\begin{equation*}
\hat{\psi} (\boldsymbol \varepsilon,s)  =  \psi^{e}(\boldsymbol \varepsilon,s) +  G_c \gamma(s, \nabla s).
\end{equation*}
The functional $\gamma (s, \nabla s)$ is assumed to be a convex function of $s$ and its gradient $\nabla s$ and can be written, in agreement with the following expressions characterizing the AT1 and the AT2 models,  respectively, as:
\begin{equation}
\gamma(s, \nabla s)=
\begin{cases}
\dfrac{3}{8}\left(\dfrac{s}{l}+l|\nabla s |^{2}\right),\, &(\text{for the AT1 model}),\\[2mm]
\dfrac{1}{2}\left(\dfrac{s^{2}}{l} + l|\nabla s |^{2}\right),\,&(\text{for the AT2 model}),
\end{cases}\label{crackdensity}
\end{equation}
where $l$ stands for a regularisation characteristic length that can be related to the Young's modulus, the fracture toughness, and the tensile strength of the material, 
as specified in Section \ref{sezion4}. 

\noindent To avoid the development of damage in compression, so to allow fracture growth only under tensile stress states, the following \lq tensile/compressive' decomposition is herein assumed for the energy density in the bulk $\psi^{e}(\boldsymbol \varepsilon,s)$ \cite{Miehe2010,  Ulmer2013, Hofacker2013, Miehe2015a, Miehe2015b, yvo1, yvo2, yvo3} and included in both AT1 and AT2 formulations: 
\begin{equation}
\psi^{e}(\boldsymbol \varepsilon,s) = g(s) \psi^{e}_{+}(\boldsymbol \varepsilon) + \psi^{e}_{-}(\boldsymbol \varepsilon),
\end{equation}
where 
$g(s)$ is a damage function that is assumed in the simple form $g(s) = \left(  1 - s \right)^{2}+k$,
where $k$ is a residual stiffness (introduced to avoid ill-conditioning) and 
the positive and negative parts of the energy density are defined as 
\begin{equation}
\label{pere}
\psi^{e}_{\pm}(\boldsymbol \varepsilon)= \dfrac{\lambda}{2} \left(\text{tr} \boldsymbol \varepsilon_{\pm }  \right)^2 + \mu  \text{tr} \left(\boldsymbol \varepsilon_{\pm}^{2} \right) ,
\end{equation}
where $\lambda$ and $\mu$ are the Lam\'e constants, 
$\text{tr} (\cdot)$ denotes the trace operator and the positive and negative parts of the strain $\boldsymbol \varepsilon_{\pm }$ are defined as follows. With reference to the spectral 
representation for the strain (with eigenvalues 
$\epsilon_i$ and unit eigenvectors $\boldsymbol e_i$), 
denoted as 
\begin{equation}
\boldsymbol \varepsilon = \sum_{i=1}^3 \epsilon_i \boldsymbol e_i \otimes \boldsymbol e_i, 
\end{equation}
the strain is additively decomposed 
as  $\boldsymbol \varepsilon = \boldsymbol \varepsilon_{+} + \boldsymbol \varepsilon_{-}$,
so that the tensile and compressive parts associated to $\boldsymbol{\varepsilon}$ are
\begin{equation}
\boldsymbol \varepsilon_+ = \sum_{i=1}^3 \langle \epsilon_i \rangle \boldsymbol e_i \otimes \boldsymbol e_i, ~~~~~~ \mbox{ and } ~~~~~~
\boldsymbol \varepsilon_- = -\sum_{i=1}^3 \langle -\epsilon_i \rangle \boldsymbol e_i \otimes \boldsymbol e_i,
\end{equation}
respectively, where the Macaulay bracket operator is defined for every scalar $x$ as $\langle x  \rangle = (x + | x |)/2$. 

\noindent A standard derivation \cite{Coleman1963} leads \eqref{pere} to the Cauchy stress tensor from the strain energy density:

     \begin{equation}
     \label{stress}
\boldsymbol \sigma  = g(s) \boldsymbol \sigma_{+} + \boldsymbol \sigma_{-}=  \{ \left(  1 - s \right)^{2} + k \} \left(  \lambda  \text{tr} \boldsymbol \varepsilon_{+}  \, \mathbf{I}  + 2 \mu   \boldsymbol \varepsilon_{+} \right) +  \lambda  \text{tr} \boldsymbol \varepsilon_{-}  \, \mathbf{I}  + 2 \mu   \boldsymbol \varepsilon_{-},
       \end{equation}
where $\mathbf{I}$ denotes the second-order identity tensor.
The thermodynamic consistency of the above constitutive theory, in agreement with the Clausius-Duhem 
inequality, has been addressed in \cite{Miehe2010}.

\subsection{Weak form of the variational problem}
\label{Variational}

The weak form corresponding to the phase field model for brittle fracture can be derived following a standard Galerkin procedure. In particular, the weak form of the coupled displacement and phase field damage problem according to Eq.\eqref{regular} is:

\begin{equation}
\delta \Pi=
\begin{dcases}
  \!\begin{multlined}[t]
  \int_{\Omega  } \boldsymbol \sigma ( \mathbf{u}) : \boldsymbol \varepsilon( \mathbf{v})    \ \mathrm{d}  \mathbf{x}  -  \int_{\Omega  }  2 \psi^e_+ (\boldsymbol \varepsilon) (1- s) \phi    \ \mathrm{d}  \mathbf{x}  + \\
\quad \quad  \quad \quad \quad \quad \quad +\int_{\Omega  } \dfrac{3}{8}G_c \ \Big\{ \frac{1}{l} \phi +  2 l \nabla  s \cdot \nabla \phi   \Big\}  \ \mathrm{d}  \mathbf{x}+ \delta \Pi_{\text{ext}}\; \quad  \text{(AT1)},
  \end{multlined}
\\[2ex]
 \!\begin{multlined}[t]
  \int_{\Omega  } \boldsymbol \sigma ( \mathbf{u}) : \boldsymbol \varepsilon( \mathbf{v})    \ \mathrm{d}  \mathbf{x}  -  \int_{\Omega  }  2 \psi^e_+ (\boldsymbol \varepsilon) (1- s) \phi    \ \mathrm{d}  \mathbf{x}  + \\
\quad \quad  \quad \quad \quad \quad \quad +
\int_{\Omega  } G_c \ \Big\{ \frac{1}{l}  s  \phi +   l\nabla  s \cdot \nabla \phi   \Big\}  \ \mathrm{d}  \mathbf{x}+ \delta \Pi_{\text{ext}}\;\; \quad  \text{(AT2)},
  \end{multlined}
\end{dcases}
\label{var1}
\end{equation}
where $\mathbf{v}$ is the vector of the displacement test functions defined on $\textbf{H}^{1}_0(\Omega)$, $\phi$ stands for the phase field test function defined on $\textrm{H}^{1}_0(\Omega)$. Eq.\eqref{var1} holds for every test functions $\mathbf{v}$ and $\phi$. The external contribution to the variation of the bulk functional in Eq.~\eqref{var1} is defined as follows:
\begin{equation}
\label{var2}
\delta \Pi_{\text{ext}} (\mathbf{u}, \mathbf{v}) =  \int_{ \partial \Omega}   \mathbf{\overline{T}} \cdot \mathbf{v} \ \mathrm{d} \Gamma +
  \int_{\Omega}   \mathbf{b} \cdot \mathbf{v}   \ \mathrm{d}  \mathbf{x}.
\end{equation}

\subsection{Finite element formulation}
\label{FE}
\noindent The mechanical problem can be stated as: given the prescribed loading condition $\mathbf{\overline{u}}_n$ and $\mathbf{\overline{T}}_n$ at step $n$, find $\mathbf{u} \in \mathbf{U}  = \left\{\mathbf{u}\, | \,   \mathbf{u} = \overline{\mathbf{u}}_n \text { on }  \partial \Omega_{u} , \mathbf{u} \in \mathbf{H}^{1}(\Omega) \right\}$ such that
\begin{equation}
\label{varu}
\mathcal{E}_{\mathbf{u}}(\mathbf{u},s; \mathbf{v}):=\int_{\Omega  } \boldsymbol \sigma(\mathbf {u}) : \boldsymbol \varepsilon( \mathbf{v})    \ \mathrm{d}  \mathbf{x}  - \int_{ \partial \Omega}   \mathbf{\overline{T}}_n \cdot \mathbf{v} \ \mathrm{d} \Gamma -
  \int_{\Omega}   \mathbf{b} \cdot \mathbf{v} \ \mathrm{d}  \mathbf{x}=0, \ \forall \mathbf{v} \in \mathbf{H}^{1}_0(\Omega),
\end{equation}
while the phase field problem is formulated as: find $s \in S  = \left\{ s \, | \, s = 0 \text { on }  \Gamma , s \in \mathrm{H}^{1}(\Omega) \right\}$ such that $\forall \phi \in \mathrm{H}^1_0(\Omega)$:
\begin{subequations}
\begin{flalign}
 \mathcal{E}^{(1)}_{s}(\mathbf{u},s; \phi)&:=\int_{\Omega  } \dfrac{3G_c l}{4}   \nabla  s \cdot \nabla \phi \ \mathrm{d}  \mathbf{x} + \int_{\Omega} 2 \psi^e_+(\boldsymbol \varepsilon) s \phi \ \mathrm{d}  \mathbf{x}+\nonumber &\\
&-\tau \int_{\Omega} \langle s_{n-1}-s  \rangle \phi \  \mathrm{d}  \mathbf{x} +\int_{\Omega} \left(\dfrac{3G_c}{8l} - 2 \psi^e_+(\boldsymbol \varepsilon)  \right)\phi\mathrm{d}  \mathbf{x}=0 , (\text{AT1}), \label{varphi1}
\end{flalign}
\begin{flalign}
\mathcal{E}^{(2)}_{s}(\mathbf{u},s; \phi):=\int_{\Omega  } G_c l   \nabla  s \cdot \nabla \phi  \  \mathrm{d}  \mathbf{x} & + \int_{\Omega} \left( \dfrac{G_c}{l} + 2 H \right) s \phi \ \mathrm{d}  \mathbf{x}+\nonumber &\\
&\quad \quad \quad \quad \quad -\int_{\Omega} 2 H \phi \ \mathrm{d}  \mathbf{x}=0,\ \ \ \text{(AT2)},  \label{varphi2}
\end{flalign}
\end{subequations}
where $H( \boldsymbol \varepsilon)=\text{max}_{\tau \in [0,t]}\left\lbrace \psi^e_+(\boldsymbol \varepsilon(\tau)) \right\rbrace$ is the strain history function, accounting for the irreversibility of crack formation \cite{Miehe2010,Msekh2015}. Notice that in (AT1) model the strain history $H$ cannot be used with the spectral model \cite{Gerasimov, Ambati2015}; to enforce irreversibility in this case the penalty term depending on $\tau >> 1$ has been introduced, where $s_{n-1}$ stands for the solution of the phase field found at the previous loading step $n-1$. 

To solve the quasi-static evolution problems for brittle fracture, isoparametric
linear triangular finite elements are used for the spatial discretization, and a staggered solution scheme is considered. Staggered schemes based on alternate minimization exploit the convexity of the energy functional with respect to each individual variable $\mathbf{u}$ and $s$ \cite{wambacq2021}. Here, an 
{\it ad hoc} developed solver has been implemented in the software FEniCS, see Alg. 1 for the algorithm description. A series of benchmark tests taken from \cite{Ambati2014,Miehe2010} has been carried out to validate the methodology.

\subsection{Newton-Raphson procedure}
\noindent Even if the mechanical problem has been split into Eqs.~\eqref{varu} and \eqref{varphi1} for AT1, or \eqref{varphi2} for AT2, so that the phase field is reduced to a linear problem, nonlinearity still remains, because of the piece-wise linearity of the constitutive law, which includes a spectral decomposition of the strain. Therefore, a consistent linearization is required, so that  the linear form defined by the residual can be written as:
\begin{flalign}
F_{\mathbf{u}}(\mathbf{u}, s ;\mathbf{v})=&\int_{\Omega  } \left\lbrace ( (1-s)^2+k ) \boldsymbol \sigma_{+}(\mathbf {u}) : \boldsymbol \varepsilon( \mathbf{v}) + \boldsymbol \sigma_{-}(\mathbf {u}) : \boldsymbol \varepsilon( \mathbf{v})  \right\rbrace \ \mathrm{d}  \mathbf{x}+\nonumber &\\  
& \quad \quad \quad \quad \quad \quad \quad \quad \quad \quad \quad \quad \quad \quad - \int_{ \partial \Omega}   \mathbf{\overline{T}} \cdot \mathbf{v} \ \mathrm{d} \Gamma -
  \int_{\Omega}   \mathbf{b} \cdot \mathbf{v}   \ \mathrm{d}  \mathbf{x}.
\end{flalign}
Given $\mathbf{u}^k$ the current Newton-Raphson approximate solution at iteration $k$, the correction $\delta \mathbf{u}$ is therefore the solution of the following linear variational problem: find $\delta \mathbf{u} \in \mathbf{U}_0  = \left\{\mathbf{u}\, | \,   \mathbf{u} = \mathbf{0} \text { on }  \partial \Omega_{u} , \mathbf{u} \in \mathbf{H}^{1}(\Omega) \right\}$ such that
$
J_{\mathbf{u}}(\delta \mathbf{u}, \mathbf{u}^k,s ; \mathbf{v})=-F_{\mathbf{u}}(\mathbf{u}^k, s; \mathbf{v}), 
$ $\ \forall \mathbf{v} \in \mathbf{H}^{1}_0(\Omega) \ ,$
and then iterate as $\mathbf{u}^{k+1}=\mathbf{u}^k + \delta \mathbf{u}$.
The Jacobian entering the formulation is 
\begin{flalign}
J_{\mathbf{u}}(\delta \mathbf{u},\mathbf{u},s ; \mathbf{v} )=\int_{\Omega  } \left\lbrace ( (1-s)^2+k ) \partial \boldsymbol \sigma_{+}(\delta \mathbf{u}, \mathbf {u}) : \boldsymbol \varepsilon( \mathbf{v}) + \partial \boldsymbol \sigma_{-}(\delta \mathbf {u}, \mathbf{u}) : \boldsymbol \varepsilon( \mathbf{v})  \right\rbrace \ \mathrm{d}  \mathbf{x}\ ;&&
\end{flalign}
for details about the terms $\partial \boldsymbol \sigma_{-}$, $\partial \boldsymbol \sigma_{+}$ we refer to \cite{Jod2020}.

\begin{algorithm}
\caption{Staggered iterative scheme for phase field fracture at a step $n \geq 1$}\label{alg:cap}
\begin{algorithmic}[1]
\State  \textbf{Input:} Displacements and phase field $(\mathbf{u}_{n-1}, s_{n-1})$ and prescribed loads $(\mathbf{\overline{u}}_{n} , \mathbf{\overline{T}}_{n})$: 
\State Initialize $(\mathbf{u}^0, s^0):=(\mathbf{u}_{n-1}, s_{n-1})$;
\For{$k \geq 1$ \ \text{staggered iteration}}: 
\State Given $s^{k-1}$, solve the mechanical problem (13): 
$\mathcal{E}_{\mathbf{u}}(\mathbf{u}, s^{k-1}; \mathbf{v})=0$ for $\mathbf{u}$, set $\mathbf{u}:=\mathbf{u}^k$;
\State Given $\mathbf{u}^k$, solve the phase field problem AT1 (14a) or AT2 (14b): 
$\mathcal{E}^{(i)}_{s}(\mathbf{u}^k, s; \phi)=0$ for $s$, set $s:=s^k$;
\If {$\text{max}\{ || \mathbf{u}^k-\mathbf{u}^{k-1} ||/||\mathbf{u}^k ||, | s^k-s^{k-1} |/|s^k| \} < \text{tol}$:}
\State set $(\mathbf{u}^k,s^k):=(\mathbf{u}_n, s_n)$;
\Else \ $k+1 \rightarrow k$.
\EndIf
\EndFor
\State  \textbf{Output:} $(\mathbf{u}_n, s_n)$.
\end{algorithmic}
\end{algorithm}

\section{Photoelastic experiments on crack trajectories}\label{PMMA:Charct}

A series of experiments: (i.) modified compact tension tests, (ii.) four-point bending tests, and (iii.) compression of samples containing a circular hole with sharp cuts at the upper and lower edges, have been designed and performed at the \lq Instabilities Lab' of the University of Trento. In such experiments, PMMA plates,  perforated with one or two circular holes,  
and containing one or two notches, have been subjected to increasing tensile opening far-field displacements, so to induce curvilinear crack growth. Notice that in all cases examined 
the curvature of the fracture follows from the presence of the voids, so that rectilinear Mode I crack growth would occur in the absence of the latter.
The tests have been conducted 
using a linear and circular  polariscope (designed and manufactured at the \lq Instabilities Lab' of the University of Trento, with quarterwave retarders for 560~nm, dark field arrangement and equipped with a white and sodium vapor lightbox at $\lambda$ = 589.3~nm, from Tiedemann \& Betz GmbH \& Co) at white light. 
Loading on the samples has been provided with an electromechanical universal testing machine (ELE Tritest 50, ELE International Ltd), equipped with a TH-KN50kN (Gefran) loading cell. 
Displacements were measured with a displacement transducer mounted internally to the testing machine. During the execution of the experiments, load and displacement were recorded with a NI cDaq acquisition system (National Instruments). Videos of the experiments (visible in the complementary supporting material) have been recorded  with a Sony PXW-FS7 camera, equipped with a Sony Fe 100-400mm F/4.5-5.6 Gm Oss lens. Experiments have been performed by imposing a constant actuation speed of 0.02~mm/s. Photoelasticity has been found to provide an outstanding capability to visualize the stress field near the crack tip and so to highlight the process of  crack growth.
In all experiments containing a notch, this was sharpened manually with a blade, to facilitate crack growth at low stress, a factor slowing down fracture propagation and facilitating its analysis. It can be anticipated  that it was  impossible to control the sharpening. Consequently, the stress for crack propagation resulted to be influenced by a random factor and thus fracture nucleation occurred at various level of applied loading.

\begin{figure}[H]
 \centering
 \includegraphics[width=1\linewidth]{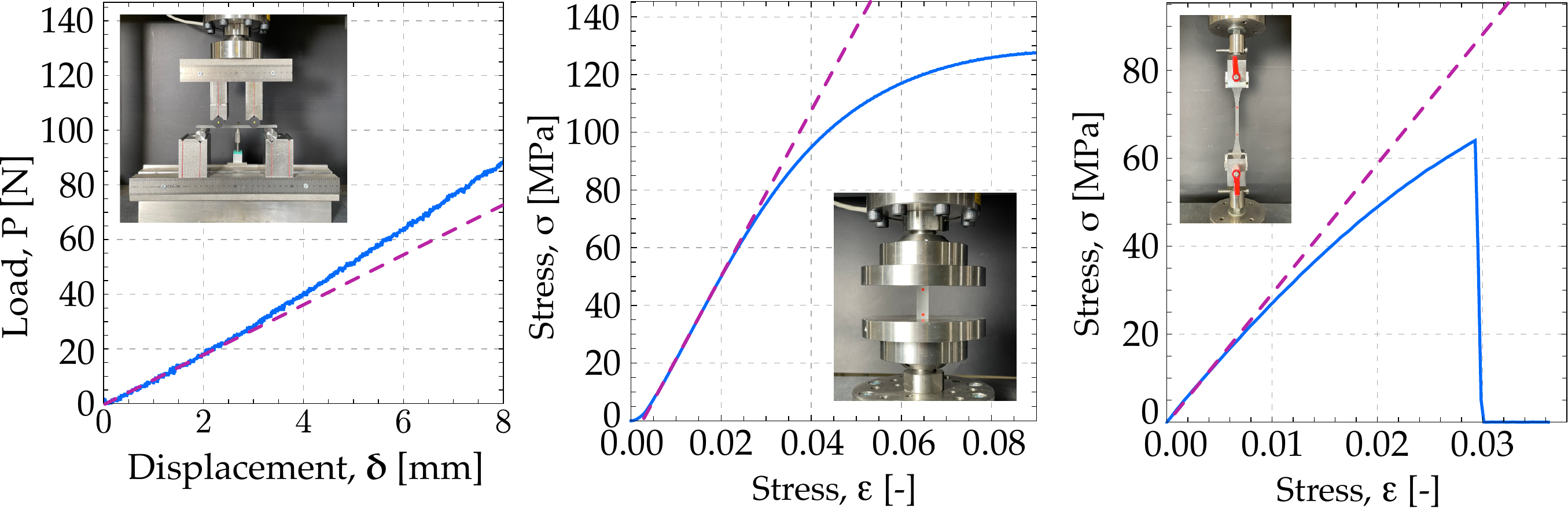}
 \caption{Example of a load/displacement curve recorded during the execution of four-point bending tests (left). 
 Examples of a stress/strain curve recorded during the execution of compression tests (center) and tensile tests (right).  Purple-dashed lines indicate the interpolating straight line used to determine the tangent modulus of elasticity, while the blue lines show the experimental data. In each graph, the insect depicts the setup used to perform the respective mechanical test.}
 \label{mech:charact}
\end{figure}

Preliminary experiments have been performed to estimate the modulus of elasticity $E$ of the PMMA used in the photoelastic tests, while the Poisson's ratio $\nu$ has been assumed equal to 0.36; from these values the Lamé constants $\lambda$ and $\mu$ required for numerical simulations have been derived.
Three different types of reference experiments (5 nominally identical samples for each type of experiment) have been carried out: uniaxial compression, four-point bending, and tensile tests, as shown in Fig.~\ref{mech:charact}.
Sample dimensions and speed of testing have been chosen accordingly with international ASTM standards. The compression tests have been performed on three prism samples (12.7x12.7x50.8~mm$^3$) by imposing a constant speed of testing of 1.3 mm/min (ASTMD-695). The four-point bending tests have been performed on three rectangular bars with cross-section 4x13~mm$^2$ and length 155 mm (support span 128~mm) by setting the speed of testing at a rate of crosshead movement of 1~mm/min (ASTMD-7264). The tensile tests have been carried out on three specimens with Type-I geometry at a speed of testing of 5~mm/min (ASTM-D638). The elastic modulus has been estimated for each type of test by averaging the values obtained from the 5 specimens. In particular, a modulus of elasticity of 2873.8$\pm$18~MPa, 3049.7$\pm$89~MPa, and 3098.5$\pm$66~MPa, has been estimated in the case of compression, four-point bending, and tensile tests, respectively.

\subsection{Modified compact tension tests}

Two series of experiments have been performed to carefully analyse how the presence of one or more voids can influence the crack trajectory.
\subsubsection*{V-notched plates containing one circular void}
\begin{figure}[H]
 \centering
 \includegraphics[width=0.8\linewidth]{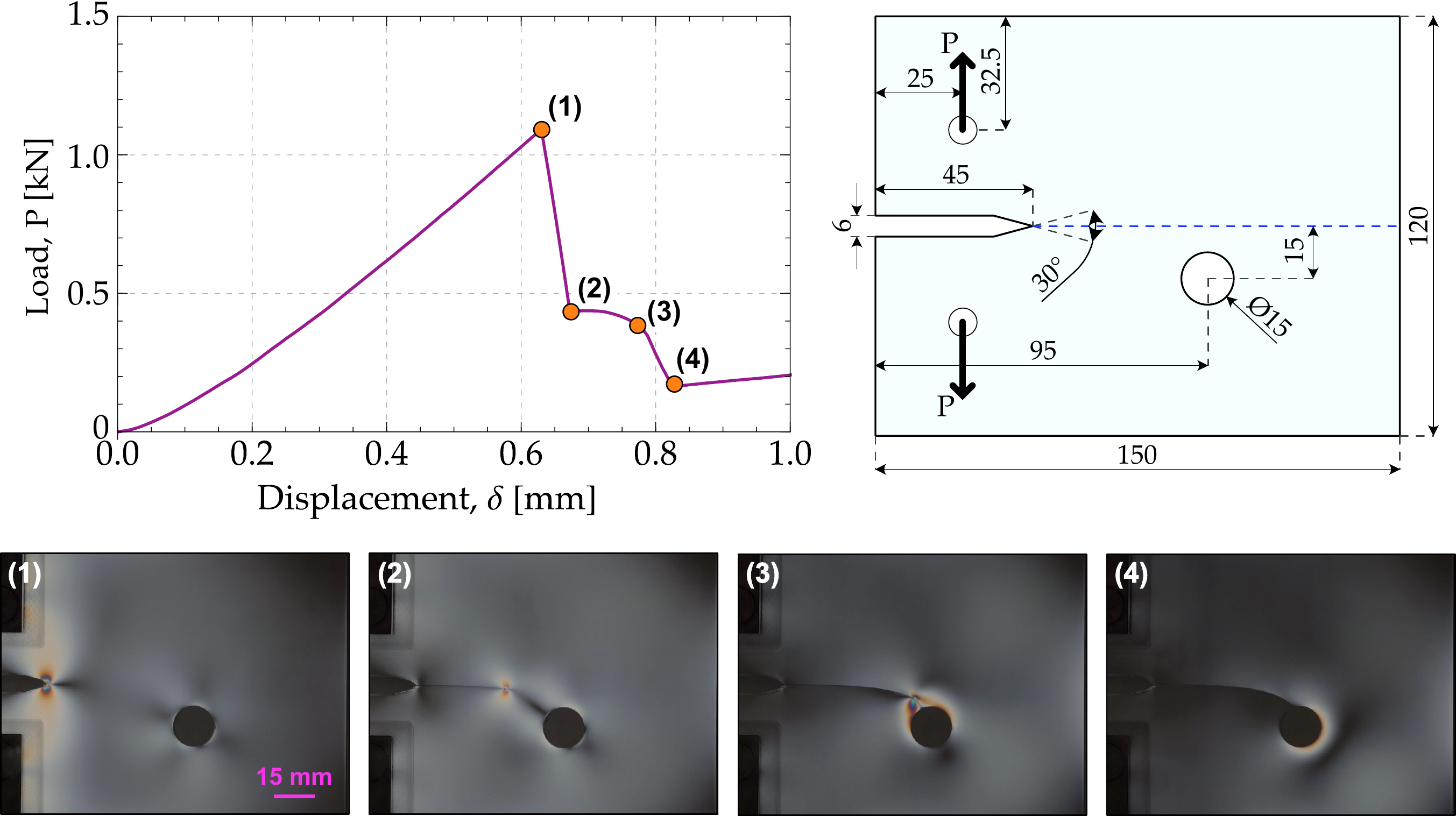}\vspace{3mm}
 \caption{Compact tension test (force/displacement graph, sample geometry with dimensions in mm and photos of the specimen) on a V-notched PMMA plate containing a circular hole. After an initial straight path, the fracture trajectory strongly deviates from rectilinearity, moving towards the hole, where its run ends.  The numbers reported in the labels of the photoelastic snapshots in the lower part of the figure correspond to the respective points highlighted on the load/displacement graph.
 }
 \label{test_1_2a}
\end{figure}

\begin{figure}[H]
 \centering
 \includegraphics[width=0.8\linewidth]{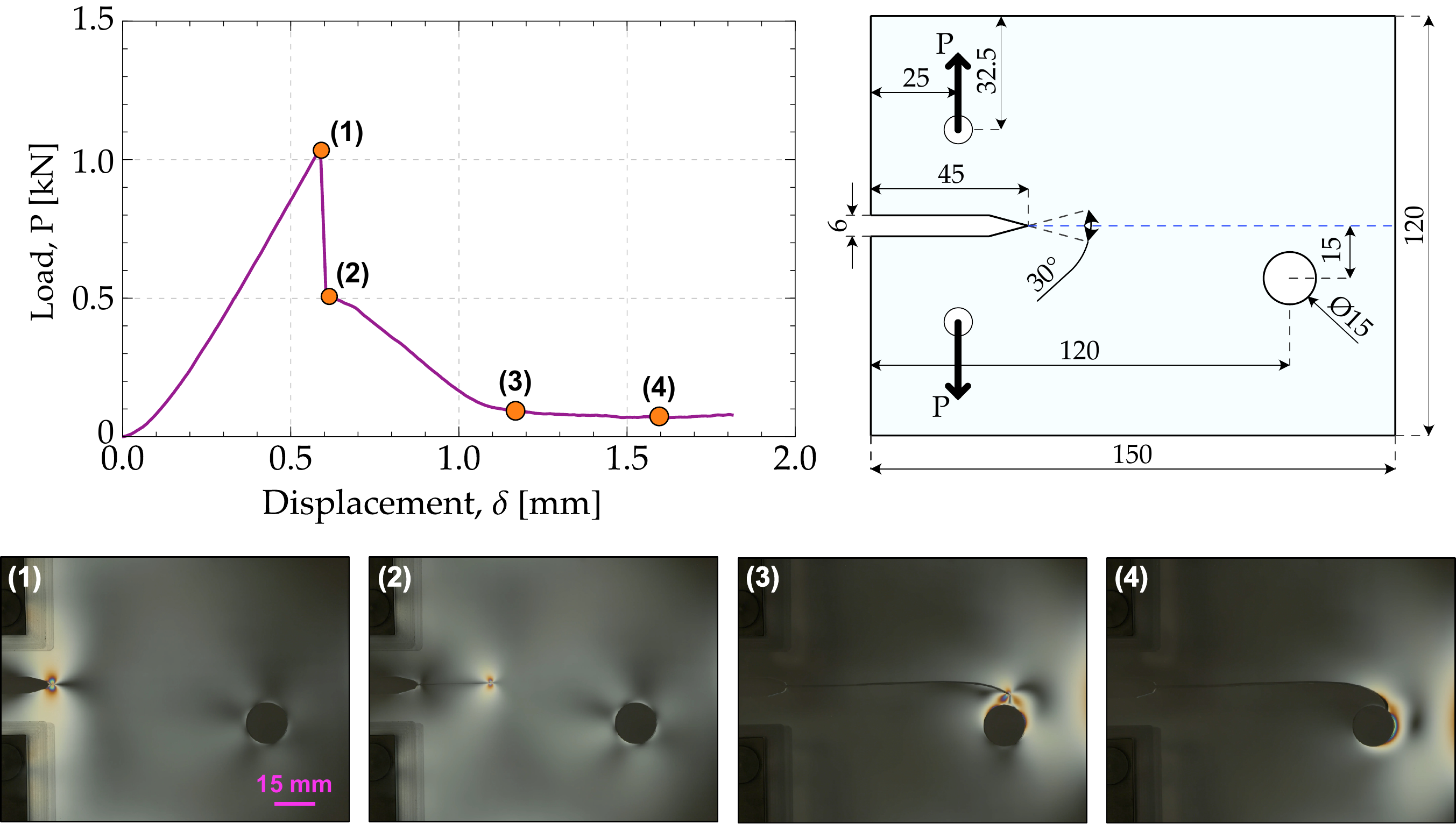}
 \caption{As for Fig. \ref{test_1_2a}, except that the 
  circular hole is located farther to the notch. The crack would be straight and horizontal in the absence of the void, which strongly deviates the fracture path. The numbers reported in the labels of the photoelastic snapshots in the lower part of the figure correspond to the respective points highlighted on the load/displacement graph.
 }
 \label{test_1_2b}
\end{figure}

Two experiments with different geometries, documenting the evolution of a crack during a compact tension test on a V-notched PMMA plate  containing a circular hole are reported in Figs. \ref{test_1_2a} and \ref{test_1_2b}. The two tests differ only in the position of the hole (both positions are illustrated in the figure, together with the exact design geometries) and 
are representative of a situation in which propagation would be rectilinear in the absence of the void. 

The load/displacement curves evidence a peak, at which crack starts propagating, and other points corresponding to an abrupt change in stiffness, marked as (1), (2), (3), and (4). Crack growth is documented with the corresponding photos. A video of experiments is available in the complementary material (movie SM1). 

In both experiments fracture initially propagates 
along a horizontal straight line, under an almost pure Mode I condition (from (1) to (2)). After this initial growth, the circular hole influences the stress state, thus inducing a 
radical change in the trajectory of the crack (3), which eventually hits the void, where it ends its run (4). 
Comparing this behaviour with the load/displacement curve, an abrupt loss of the applied driving force occurs initially (from (1) to (2)), corresponding to a fast crack propagation. 
The loss of strength between points (3) and (4) is much less pronounced in the test reported in 
Fig. \ref{test_1_2b}
than that reported in
Figs. \ref{test_1_2a}.

This effect follows from the fact that the plate becomes more compliant in the former test than in the latter, where the crack trajectory is shorter.

The crack propagation speeds referred to the test reported in Fig. \ref{test_1_2a} are, from (1) to (2) $>$0.98 m/s, from (2) to (3) 5.45$\times$10$^{-3}$ m/s, and from (3) to (4)  6.37$\times$10$^{-2}$ m/s. 
While, for the test reported in Fig. \ref{test_1_2b}, the crack speeds are: from (1) to (2) $>$1.26 m/s, from (2) to (3) 1.98$\times$10$^{-3}$ m/s, and from (3) to (4)  2.96$\times$10$^{-4}$ m/s.


\subsubsection*{V-notched plate containing two circular voids}

\begin{figure}[H]
 \centering
 \includegraphics[width=0.8\linewidth]{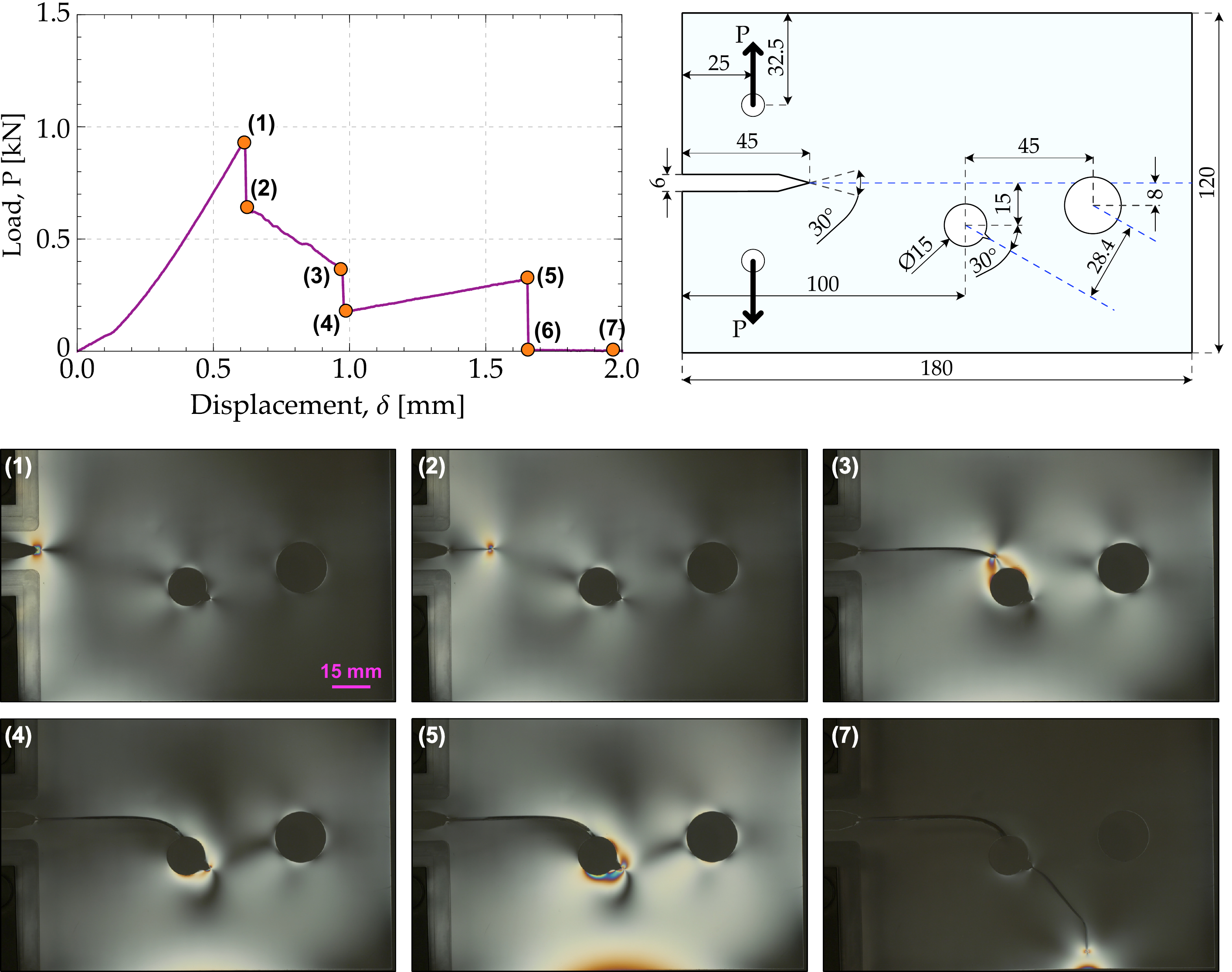}\vspace{3mm}
 \caption{
 Compact tension test (force/displacement graph, sample geometry with dimensions in mm and photos of the specimen) on a V-notched PMMA plate containing two circular holes. After an initial straight propagation, the crack hits a hole, which contains a second V-shaped notch inclined at 30$^\circ$. The latter notch induces a secondary crack growth, so that the crack terminates its run against the lower boundary of the plate. The numbers reported in the labels of the photoelastic snapshots in the lower part of the figure correspond to the respective points highlighted on the load/displacement graph.
 }
 \label{test_3_4a}
\end{figure}

\begin{figure}[H]
 \centering
 \includegraphics[width=0.8\linewidth]{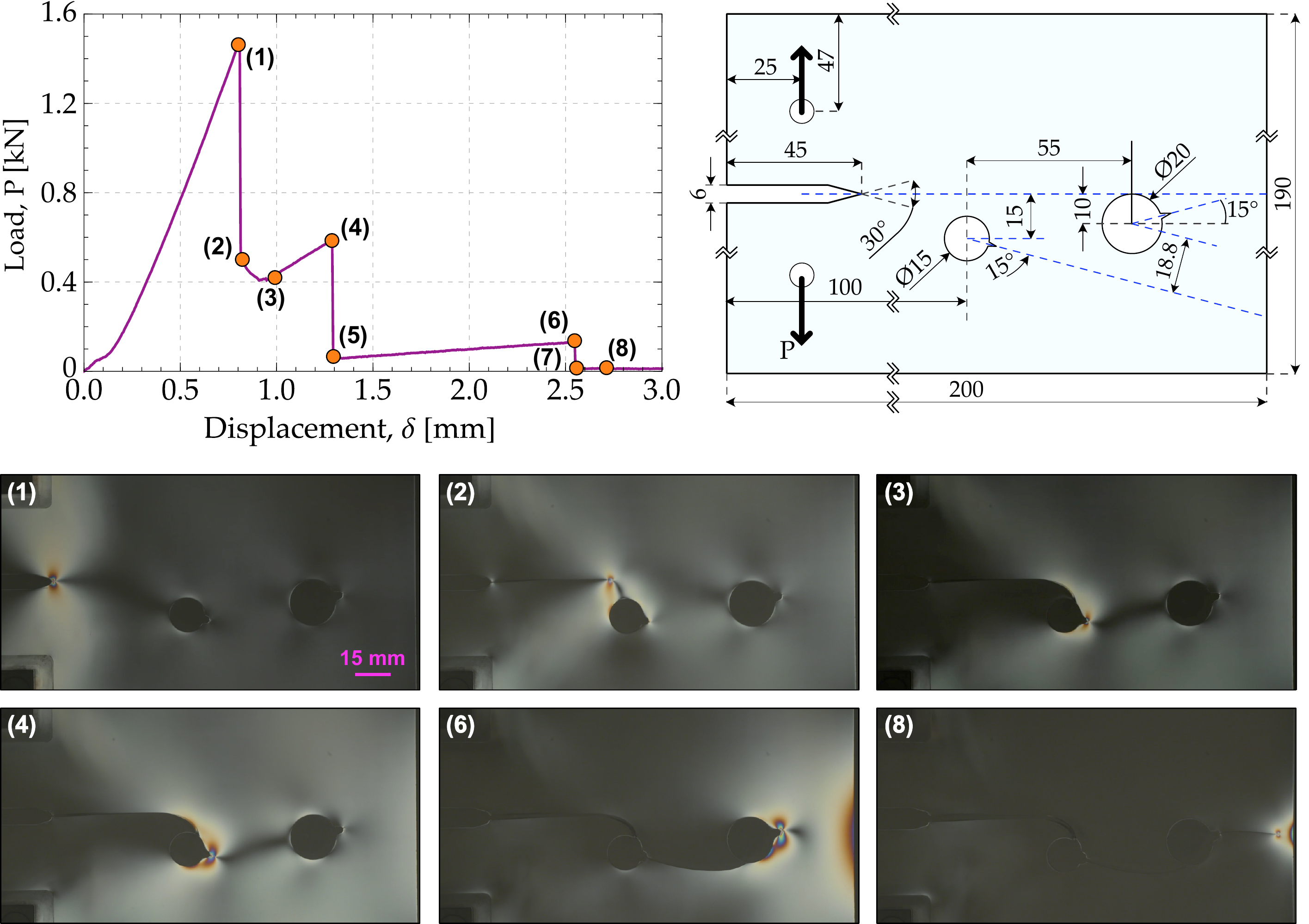}
 \caption{
 As for Fig. \ref{test_3_4a}, except that the two circular holes on the right contain a second V-shaped notch inclined at 15$^\circ$. 
 The two latter notches induce a secondary and tertiary crack growth, so that the crack hits the second void from where another notch produces a final crack propagation. The numbers reported in the labels of the photoelastic snapshots in the lower part of the figure correspond to the respective points highlighted on the load/displacement graph.
 }
 \label{test_3_4b}
\end{figure}

Two experiments referring to a 
compact tension test on a V-notched PMMA plate containing two circular holes are reported in Figs. \ref{test_3_4a} and \ref{test_3_4b}. It should be noticed that 
the hole closest to the notch contains an additional tiny V-shaped notch on its surface, that permits a 
secondary crack nucleation and growth, after this void is
hit by the crack. 
While the notch inducing the initial crack growth is horizontal, the second 
is inclined at 30$^\circ$ in Fig. \ref{test_3_4a} and at 15$^\circ$ in Fig. \ref{test_3_4b}.

Results of the tests on samples with two holes show that the orientation of the secondary V-shaped notch has an important effect on the
force/displacement curve and on 
crack trajectory. 
While in both cases the crack nucleates at the V-notch and is attracted by the first hole, 
which is hit after a high-speed propagation, the subsequent stages of propagation differ in the two samples. 
In particular, a secondary crack is nucleated in both experiments reported in 
Figs. \ref{test_3_4a} and \ref{test_3_4b}, but in the former figure  
the secondary crack it is more \lq attracted' by the lower boundary of the sample, where it terminates its run. In a different vein, the secondary crack in the other test deviates towards the second circular hole, which is 
impinged after a slow propagation at increasing load. In the second 
hole, a third notch triggers a third, and final, crack nucleation and growth.

The crack propagation speeds for the test reported in Fig. \ref{test_3_4a}  are: from (1) to (2) $>$0.82 m/s, from (2) to (3)  2.19$\times$10$^{-3}$ m/s, from (3) to (4) 2.33$\times$10$^{-2}$ m/s, from (4) to (5)  0 m/s, from (5) to (6)  $>$1.96, and from (6) to (7)  3.19$\times$10$^{-3}$. 
The propagation speeds for the test reported in Fig. \ref{test_3_4b} are: from (1) to (2) $>$2.54 m/s, from (2) to (3)  2.84$\times$10$^{-3}$ m/s, from (3) to (4)  0 m/s, from (4) to (5)  $>$2.09 m/s, from (5) to (6)  0, from (6) to (7)  $>$1.14, and from (7) to (8)  6.60$\times$10$^{-5}$.

The experiments reported in Figs. \ref{test_3_4a} and \ref{test_3_4b}
show that circular holes and V-shaped notches can be designed 
to induce crack paths with desired geometries.

The force/displacement curve evidences in both experiments several peaks, zones of abrupt load decrease and zones of low crack growth at increasing load.

In all the above reported experimental tests, the circular holes interacted with the crack, deviating its path towards 
according to the rotation of the principal stress axes.  Effects related to the complexity of the stress field  
can also be noticed at the naked eye on post-mortem samples, so that 
zones of fast crack propagation are characterized by mirror-like surfaces, while  slow crack speed is marked by high surface roughness, Fig. \ref{image_intro}. 

Surface roughness after failure was examined  using a confocal profilometer Leica DCM3D 
(available at the experimental laboratory for  Multi-scale Analysis of Materials, MUSAM-Lab,  IMT, Lucca). 
At the point where the crack deviates from the rectilinear path (typically induced by a pure Mode I stress state) and reduces its speed, the crack surface changes morphology. During the Mode I regime, the surface is characterized by a very fine random sub-microscale roughness,  invisible at the naked eye. The root mean square roughness in that region is about 80~nm. When the crack becomes curved, the stress state is inducing a surface corrugation of large amplitude, also visible at the naked eye, with a root mean square roughness value increasing up to 25~$\mu$m.

\begin{figure}[H]
 \centering
 \includegraphics[width=0.9\linewidth]{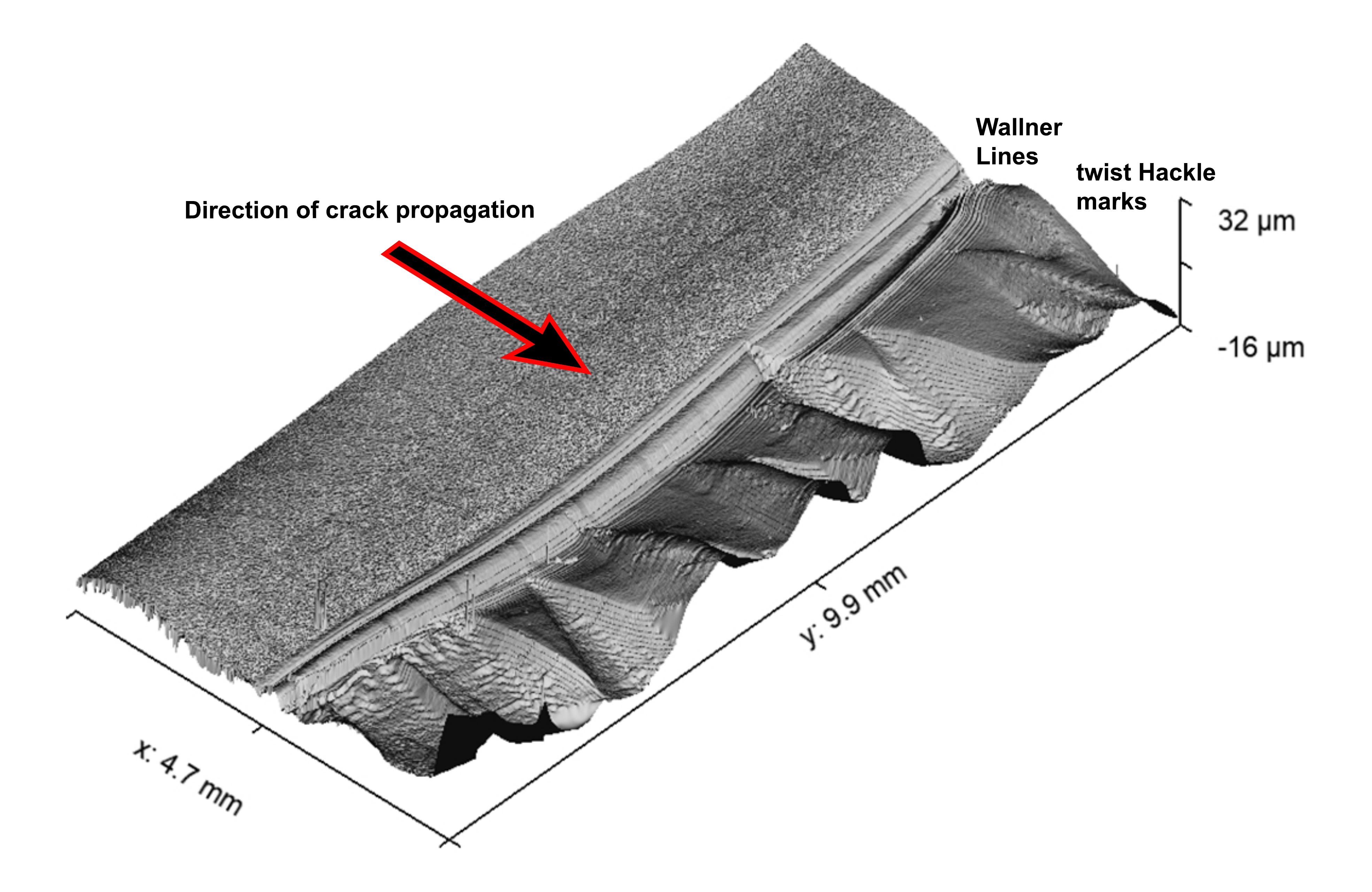}
 \caption{Surface 
 roughness after failure (examined with a confocal profilometer Leica DCM3D) shows that  
  crack morphology abruptly changes at the transition from straight (fast) to curved (slow) paths. The transition involves formation of Wallner lines, followed by twist Hackle marks.}
 \label{profilometer}
\end{figure}

\begin{figure}[H]
 \centering
 \includegraphics[width=0.6\linewidth]{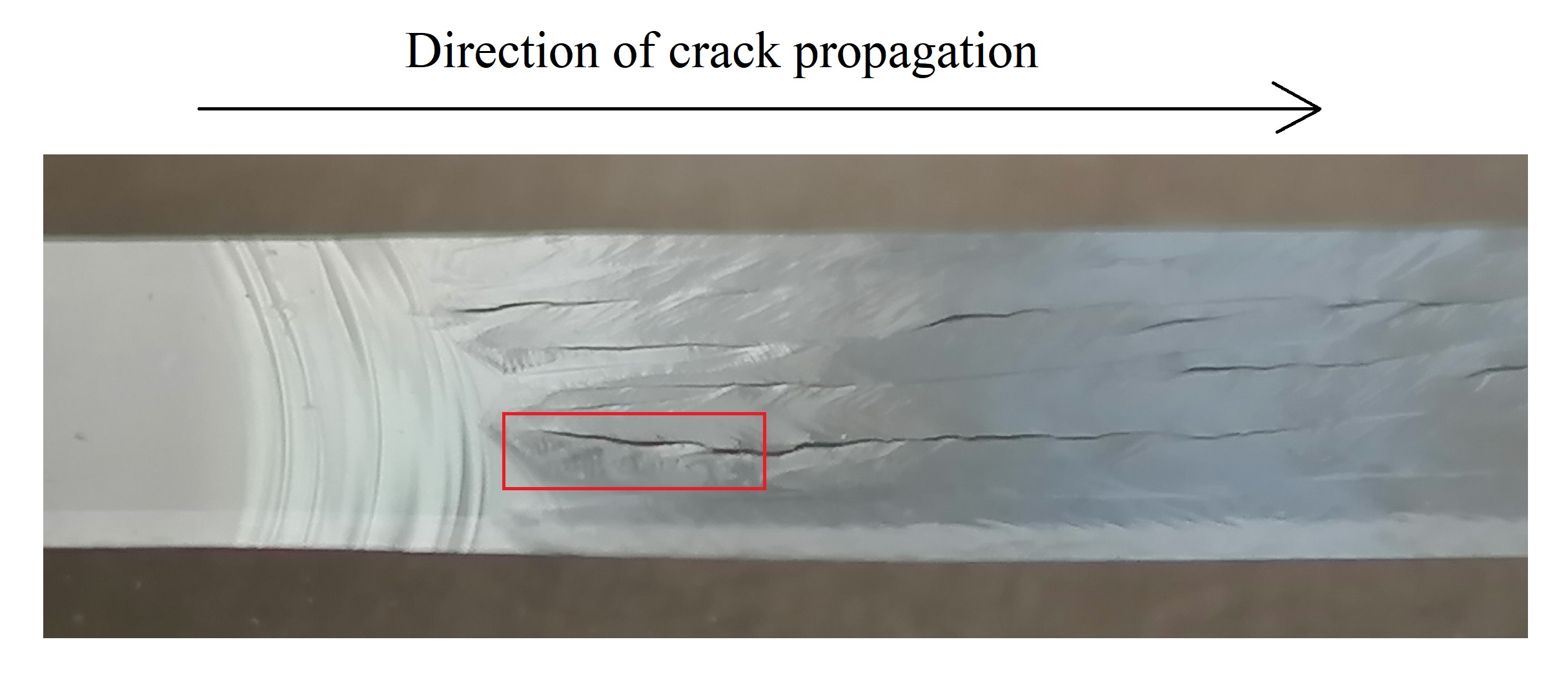}\\
 \includegraphics[width=0.7\linewidth]{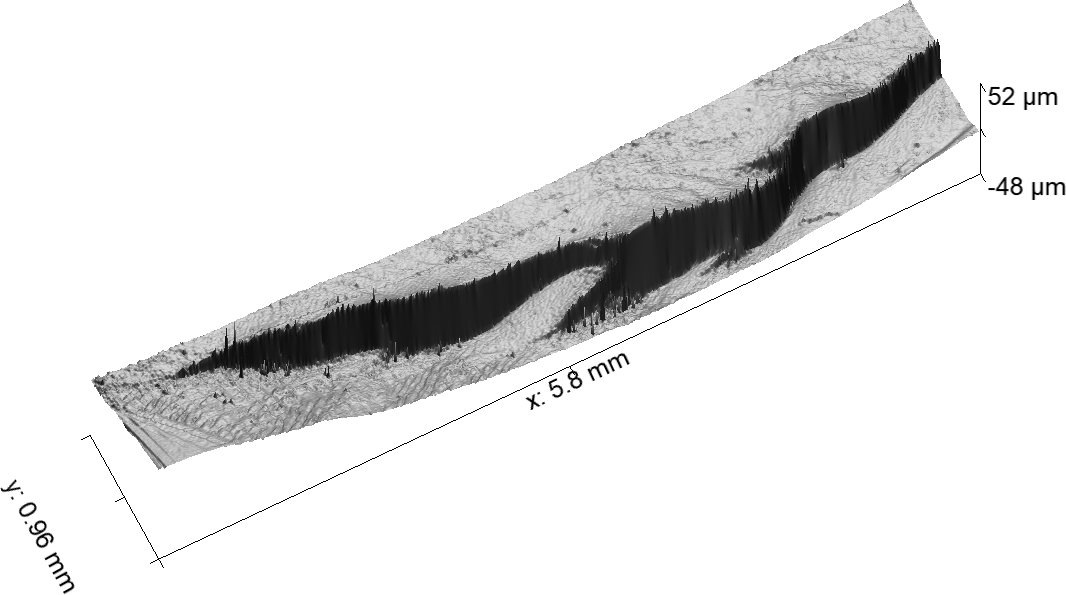}
 \caption{Longitudinal view of surface crack morphology at deviation from 
 rectilinearity (fast/slow speed transition) showing the formation of 
 Wallner lines, followed by twist Hackle marks. Upper part: photo  (taken with a confocal profilometer Leica DCM3D) of the surface; Lower part: quantitative 3D profilometric image of a 
 twist Hackle mark
 (the portion shown red in the upper part).}
 \label{profilometer2}
\end{figure}

Examples of such fast/slow crack-speed  transition are shown in Figs.~\ref{profilometer} and \ref{profilometer2}. The former figure  
(The latter figure) 
corresponds to the fracture surface at point (2) of the specimen
reported in Fig. \ref{test_1_2a} (reported in 
Fig. \ref{test_3_4a}).

In both cases at the transition, the crack initiates a curved path and, simultaneously, a wavy fine corrugation turns into remarkable undulations. 

In both cases, the change in direction and speed of the crack
is evidenced 
by  Wallner lines (the curved lines orthogonal to the direction of crack propagation, indicating that the crack has reached a terminal velocity) followed by twist Hackle marks (appearing as long scratches parallel to the direction of crack propagation, denoting a rotation of the crack-driving tensile stress)
\cite{Hull,Hayes}.

\subsection{Four-point bending tests}

\begin{figure}[H]
 \centering
 \includegraphics[width=0.9\linewidth]{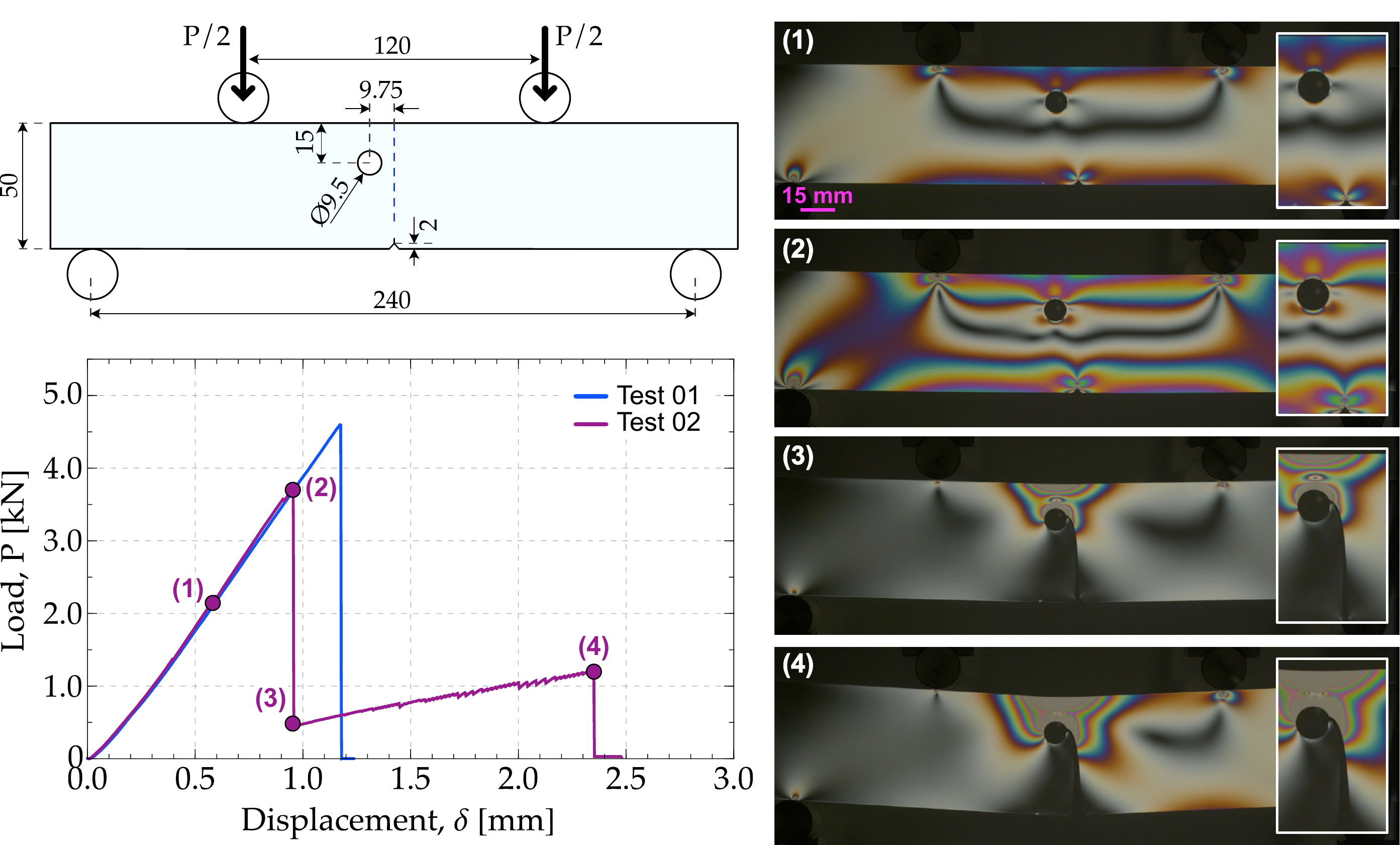}
 \vspace{-3mm}
 \caption{Four-point bending tests on a V-notched PMMA plate with  an eccentric hole. Experimental geometry (dimensions in mm), load displacement curve for two different specimens and photos of the photo-elastic response of the specimens during the test. After an initial straight growth, the fracture hits the circular void and the specimen fails abruptly.
 The differences in the peaks in the force-displacement diagram are related to the diversity 
 of the sharpening of the notch. The second peak refers to final failure of the sample. The numbers reported in the labels of the photoelastic snapshots in the right part of the figure correspond to the respective points highlighted on the load/displacement graph.
 }
 \label{exp:4PBT1}
\end{figure}

\begin{figure}[H]
 \centering
 \includegraphics[width=0.9\linewidth]{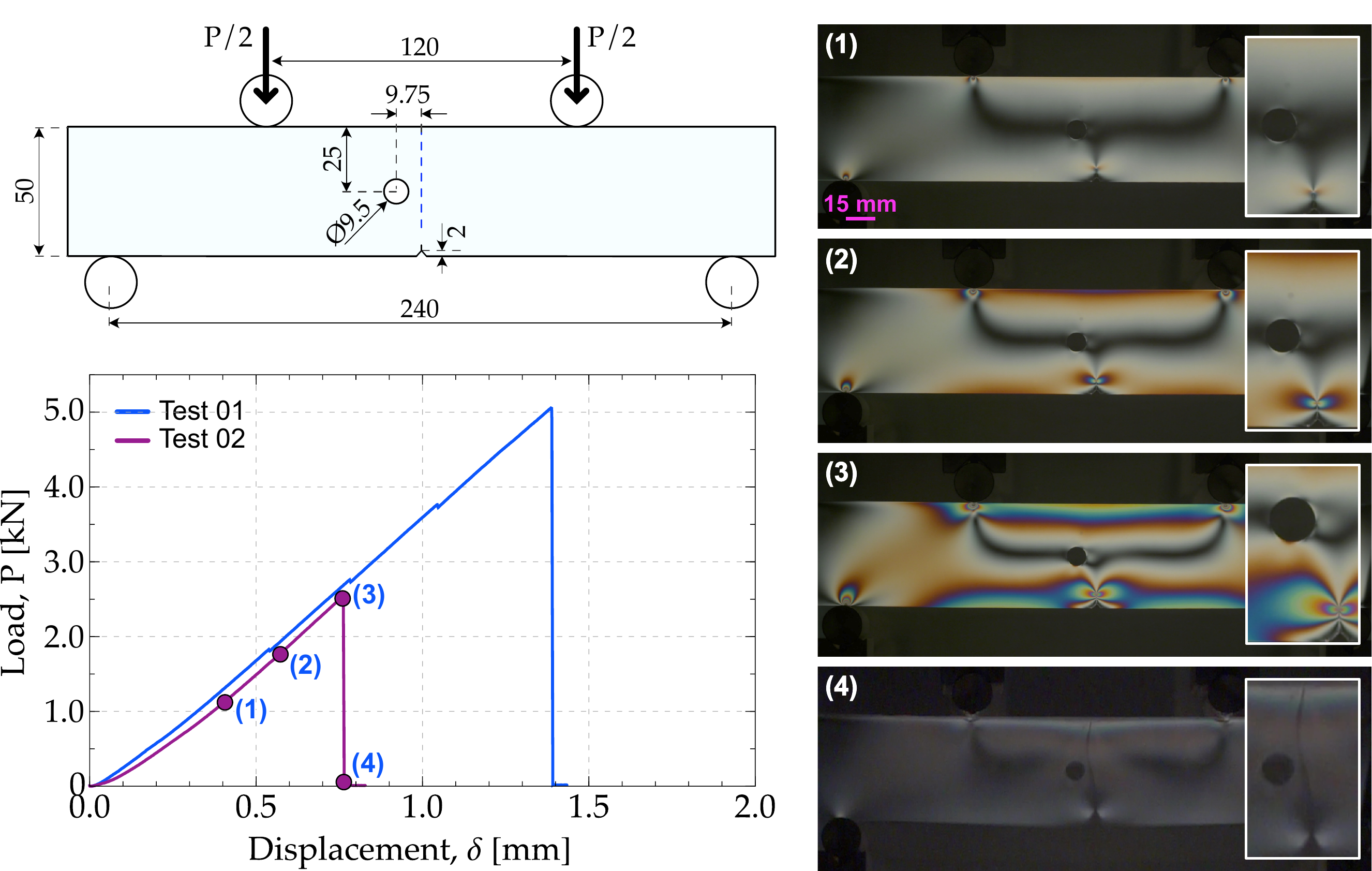}
 \vspace{-3mm}
 \caption{
 As for Fig. \ref{exp:4PBT1}, except that the crack is deviated from purely vertical trajectory towards the circular void, but the latter is not hit. The higher peak in the force/displacement curve corresponds to an experiment where the notch was not sharpened. The numbers reported in the labels of the photoelastic snapshots in the right part of the figure correspond to the respective points highlighted on the load/displacement graph.
}
 \label{exp:4PBT2}
\end{figure}

Four-point bending tests have been performed on two different pairs of PMMA specimens. The two samples forming the pairs are nominally identical, while the difference between the two pairs is in the eccentricity of a circular hole with respect to a central notch, as shown in the insets of Figs. \ref{exp:4PBT1} and \ref{exp:4PBT2}. The tests have been conducted until rupture under displacement control. A video of  experiments is available in the complementary material (movie SM2).

The V-notches were all sharpened by hand except one (Test 01 in  Fig. \ref{exp:4PBT2}). The fact that it was impossible to control the sharpness leads to a remarkable variation of the peak force recorded by the testing machine at crack nucleation (Figs.  \ref{exp:4PBT1} and \ref{exp:4PBT2}). The trajectory of the crack, visible from the photos, is initially straight and then it deviates towards the void. In the case of Fig. \ref{exp:4PBT1}, the void is impinged by the crack before final failure, while the crack is only deviated, without touching the void, in the case of Fig. \ref{exp:4PBT2}. 
 Note that the crack propagation is very fast, with a speed exceeding 20 m/s.


\subsection{Compression of samples containing a circular hole with sharp cuts at the upper and lower edges}

\begin{figure}[H]
 \centering
 \includegraphics[width=0.9\linewidth]{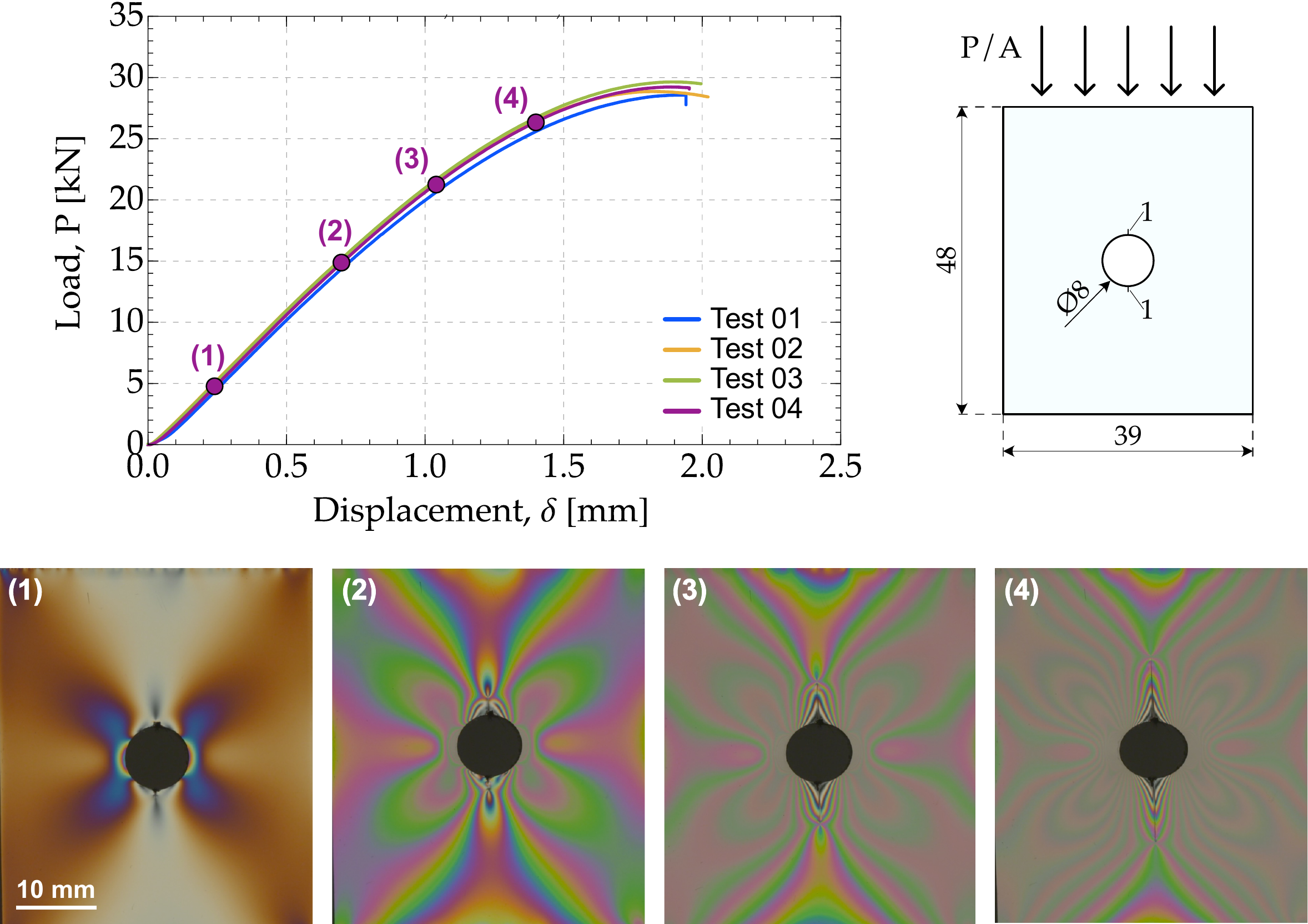}
 \vspace{-3mm}
 \caption{
 Slow vertical crack propagation from pre-cracks cut on the 
 upper and lower edge of a circular void in a plate subject to compression. The fracture develops in a straight path. 
 Load displacement curve for two different specimens (upper part on the left), experimental geometry (dimensions in mm, upper part on the right), and photos of the photo-elastic response are reported. The numbers reported in the labels of the photoelastic snapshots in the lower part of the figure correspond to the respective points highlighted on the load/displacement graph.}
 \label{exp:compression}
\end{figure}

Four final tests have been performed on nominally identical prismatic PMMA samples containing a circular void and subject to uniaxial compression, see Fig. \ref{exp:compression}. Vertical pre-cracks with depth of approximately 1~mm have been cut by hands at the upper and lower edges of the hole, to induce crack growth due to the tensile stress developing there. 
Note that in the absence of the pre-cracks, experiments not reported for brevity show that the samples fail under compression, initially forming shear bands and eventually buckling out-of-plane.
The four tests with sharp cuts have been designed to be different from all the others previously reported, because now propagation is slow and rectilinear. 

Details of the experimental set-up, with the force/displacement curves and photos at different stages of load are reported in Fig.~\ref{exp:compression}. A video of an experiment is available in the complementary material (movie SM3). 

The four samples provided almost the same results, showing that 
the influence of the sharpness of cuts is not sufficient to substantially influence the fracture growth, which resulted to be very slow. 

The propagation speeds for the Test 04 reported in the figure are: from (1) to (2) 3.21$\times$10$^{-2}$ m/s, from (2) to (3)  7.66$\times$10$^{-2}$ m/s, and from (3) to (4)  1.01$\times$10$^{-1}$ m/s.
The fracture growth is rectilinear in the vertical direction and occurs while the load grows steadily and reaches a peak when plasticity spreads and an out-of-plane buckling initiates.

\section{Phase field simulations of crack trajectories in plates with notches and holes}\label{sezion4}

This section is devoted to the assessment of the effectiveness of the AT1 and AT2 phase field models for brittle fracture, equipped with spectral decomposition, to predict complex crack trajectories observed in the experiments and resulting from a stress state ranging from pure tensile to tensile-compressive.

The presented phase-field formulation relies completely on four parameters: two elastic parameters (Lamé constants for an homogeneous isotropic elastic material, related to the engineering constants $E$ and $\nu$ reported in section~\ref{PMMA:Charct}) and the phase parameters $G_c$ (fracture energy) and $l$ (internal length-scale). All the simulations with the phase-field have been performed under the hypothesis of plane stress (2D analysis), consistently with the experimental conditions maintained during the tests. 
Although a small variation in the values of modulus of elasticity has been observed depending on the type of performed test (compression, four-point bending, or tensile, see section~\ref{PMMA:Charct}), an average value of $E$=3000~MPa has been considered in all the numerical simulations. 

\sloppy The fracture energy $G_c$ was set equal to 0.7~MPa~mm, in agreement with data available in the literature. In fact, by considering a toughness of \mbox{$K_{I,c} \in \{1.1,1.5\}$~MPa$\sqrt{\text{m}}$}, one obtains G$_c=K^2_{I,c}/E$ or G$_c \in \{0.4,0.7 \}\,\textup{MPa mm}$.
As far as the regularization parameter $l$ is concerned, two expressions are suggested in \cite{Tanne} for the AT1 and the AT2 models:
\begin{equation}
l= \left\{
\begin{array}{ll}
\dfrac{3}{8} \dfrac{\rm{G_c} E}{\sigma^2_{\rm max}}, & \text{(AT1)},\\ [5 mm]
\dfrac{27}{256} \dfrac{\rm{G_c} E}{\sigma^2_{\rm max}},  & \text{(AT2)},
\end{array}
\right. 
\label{eq:internal_length}  
\end{equation} where $\sigma_{\rm max}$ is the strength of the material under uniaxial tension. Such a simple formulation for the regularization parameter opens up to an interpretation for $l$ as a material parameter. Nonetheless, equations (\ref{eq:internal_length}) are based on severe assumptions, the most limiting of which are a null value for the Poisson ratio and homogeneity of strain and phase fields up to the instant of crack nucleation. Therefore, the above formulae can be interpreted as providing only indicative values for the internal length scale parameter. Applying Eqs. \eqref{eq:internal_length} to PMMA, $l=0.174$ mm is obtained for the AT1 model and $l=0.049$ mm for the AT2 model. Using such values, we have found that the AT1 model is able to predict a maximum force in line with the values measured in the experiments, while the AT2 model systematically predicts higher values. Therefore, for the AT2 model the length scale parameter has been increased to $l=0.1$ mm (for the modified compact test 01) and to $0.2$ mm for all the other compact tests, so that a reasonable quantitative agreement has been achieved between simulations and experimental results in terms of maximum force.

\subsection{Simulation of the modified compact tests}
 
Numerical simulations of the experiments shown in Figs. \ref{test_1_2a}--\ref{test_3_4b} on modified compact tests are now reported. 
Loading is idealized by imposing on the upper of the two little holes, machined in the sample to allow connection with the testing equipment, a displacement $u=2.5 \times 10^{-4} \ \rm mm$ subdivided into $6000$ equally-spaced loading steps. At the same time the lower of the two little holes was kept fixed. The finite element meshes used for simulation of Test 01 and Test 02 consisted of $99077$ points and $196938$ linear triangular finite elements. An \textit{ad hoc} mesh refinement was implemented where the crack was expected to nucleate/propagate, so that a mesh size of $h=0.04\ \rm mm$ was reached in that zone, a value smaller than $l/2$ \cite{Miehe2010b}. In order to compare numerical predictions with photoelastic results, the in-plane principal stress difference  $\sigma_1-\sigma_2= 2 \sqrt{\left( \dfrac{\sigma_{11}- \sigma_{22}}{2} \right)^2 + \sigma^2_{12}}$  has been computed, which correlates with the experimentally observed fringe patterns. A video of simulations is available in the complementary material (movie SM4).

\subsubsection*{V-notched plate containing one circular void}
\begin{figure}[H]
 \centering
 \includegraphics[width=\linewidth]{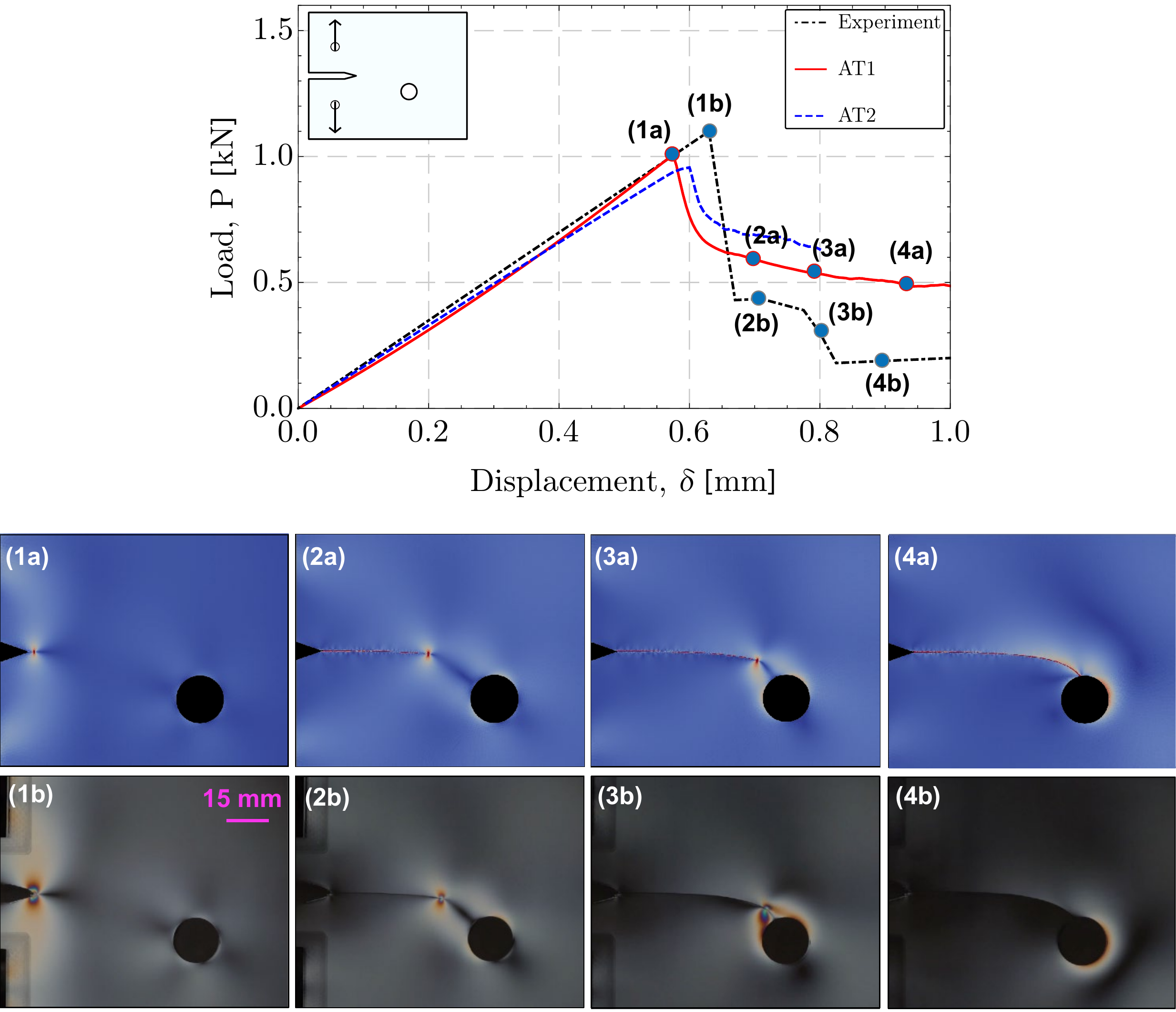}
 \caption{Comparison between simulation and experiment during the modified compact test 01 on a V-notched PMMA plate containing a circular hole. Force/displacement relation (upper part); crack path and photoelastic fringes at different displacements: simulation (central part) and experiment (lower part, from Fig. \ref{test_1_2a}). The numbers reported in the labels of the snapshots of the figure (central and lower parts) correspond to the respective points highlighted on the load/displacement graph.}
 \label{test_1_phase}
\end{figure}

\begin{figure}[H]
 \centering
 \includegraphics[width=\linewidth]{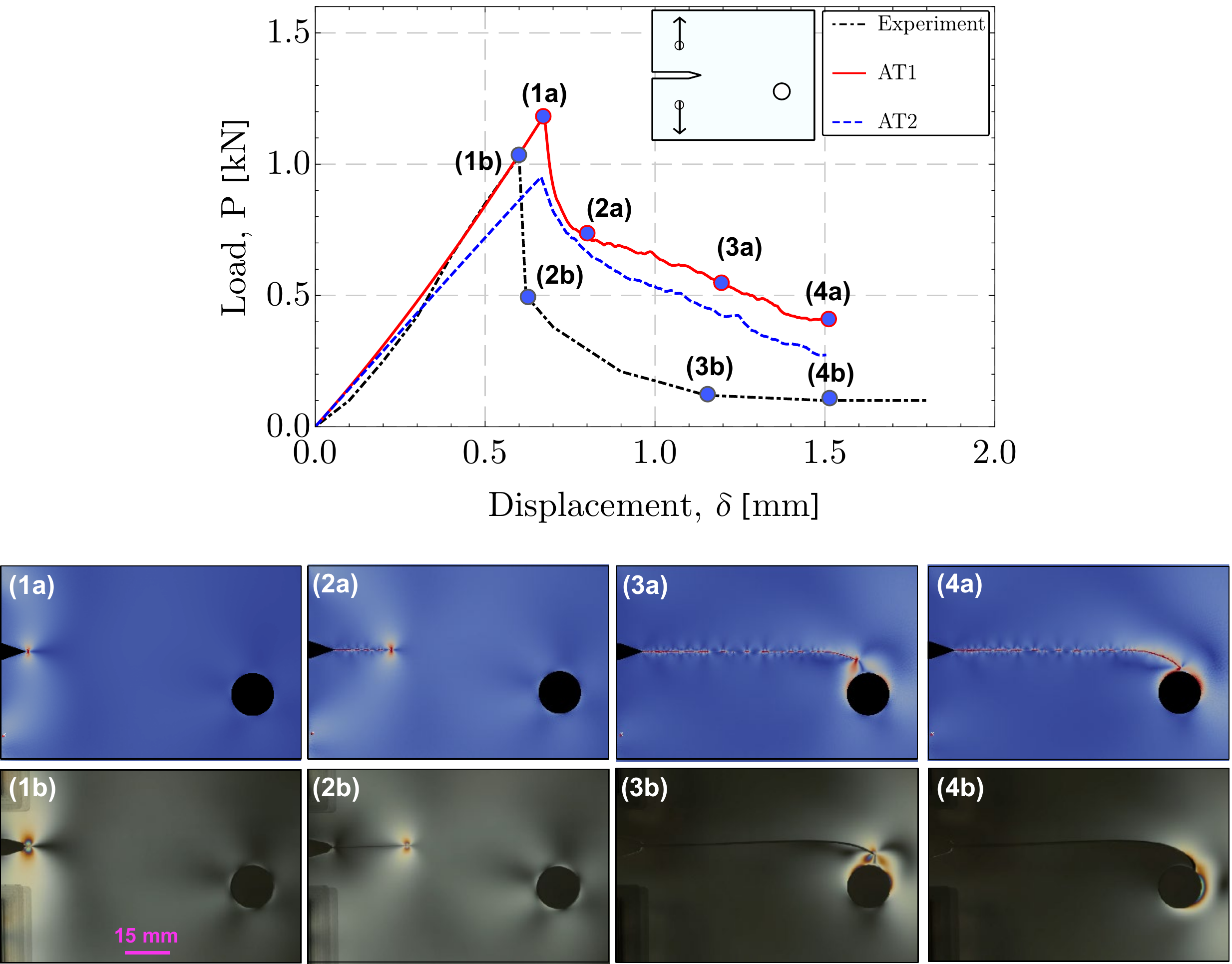}
 \vspace{-5mm}
 \caption{
 Comparison between simulation and experiment during the modified compact test 02 on a V-notched PMMA plate containing a circular hole. Force/displacement relation (upper part); crack path and photoelastic fringes at different displacements: simulation (central part) and experiment (lower part, from Fig. \ref{test_1_2b}). The numbers reported in the labels of the snapshots of the figure (central and lower parts) correspond to the respective points highlighted on the load/displacement graph.
 }
 \label{test_2_phase}
\end{figure}

The graphs on the upper part of Figures \ref{test_1_phase} and \ref{test_2_phase} show the comparison between experimental and simulated load-displacement curves for the compact tests 01 and 02, respectively. The central  part of Figures \ref{test_1_phase} and \ref{test_2_phase} present the maps of the in-plane stress difference $\sigma_{1} - \sigma_{2}$ at some characteristic step of crack growth (indicated with numbers in the figures), and are directly compared with photoelastic fringe patterns (reported in the lower part). It is clear that both AT1 and AT2 models provide similar fringe patterns and crack paths, which closely match the experimental data. The crack deviates from an initial straight path and it is attracted by the circular hole where it terminates.
Both models satisfactorily predict the peak loads observed in the experiments, while present some difficulties in reproducing the sudden drop in the load carrying capacity during the softening branch. 

Although the proposed phase field models are quasi-static, a qualitative assessment of the crack tip velocity is however attempted by computing the ratio between the crack length increment in a pseudo-time step and the corresponding time step size. Numerical results have shown for the modified compact tests 01 and 02 a trend with an increasing crack tip velocity up to the point where the crack tip tends to deviate from the straight trajectory and the stress state evolves from pure tensile to tensile-compressive. After that point, which occurs for a crack length of about 60 mm in both tests, a progressive slowing down of crack tip velocity is noticed. This trend, in qualitative agreement with the experimental observation, would deserve further quantitative analyses based on a dynamic phase field formulation \cite{Corrado}, which falls however outside the scope of the present article.    

\subsubsection*{V-notched plate containing two circular voids}
\begin{figure}[H]
 \centering
 \includegraphics[width=\linewidth]{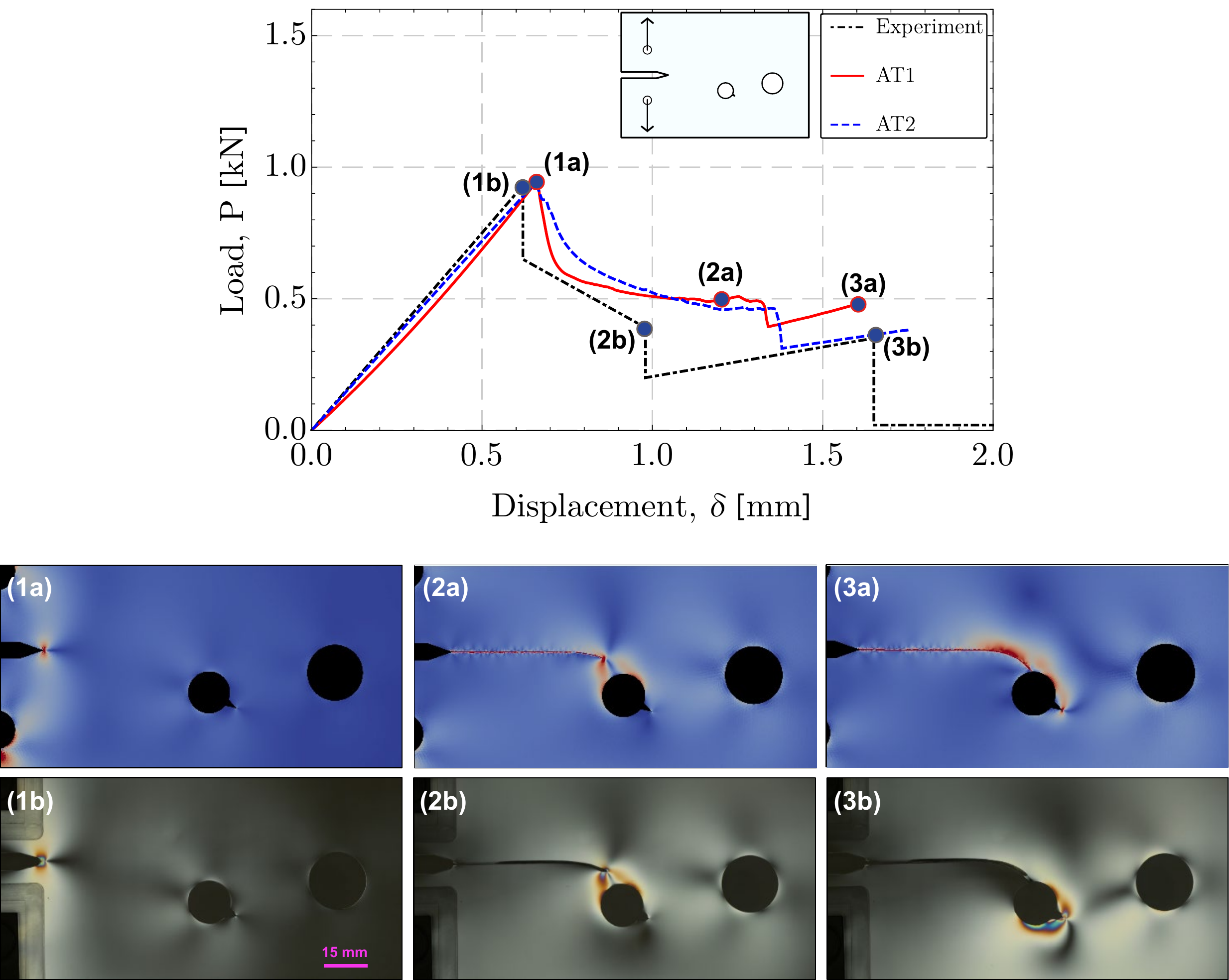}
 \vspace{-5mm}
 \caption{
 Comparison between simulation and experiment during the modified compact test 03 on a V-notched PMMA plate containing two holes. Force/displacement relation (upper part); crack path and photoelastic fringes at different displacements: simulation (central part) and experiment (lower part, from Fig. \ref{test_3_4a}). The numbers reported in the labels of the snapshots of the figure (central and lower parts) correspond to the respective points highlighted on the load/displacement graph.
 }
 \label{test_3_phase}
\end{figure}

\begin{figure}[H]
 \centering
 \includegraphics[width=\linewidth]{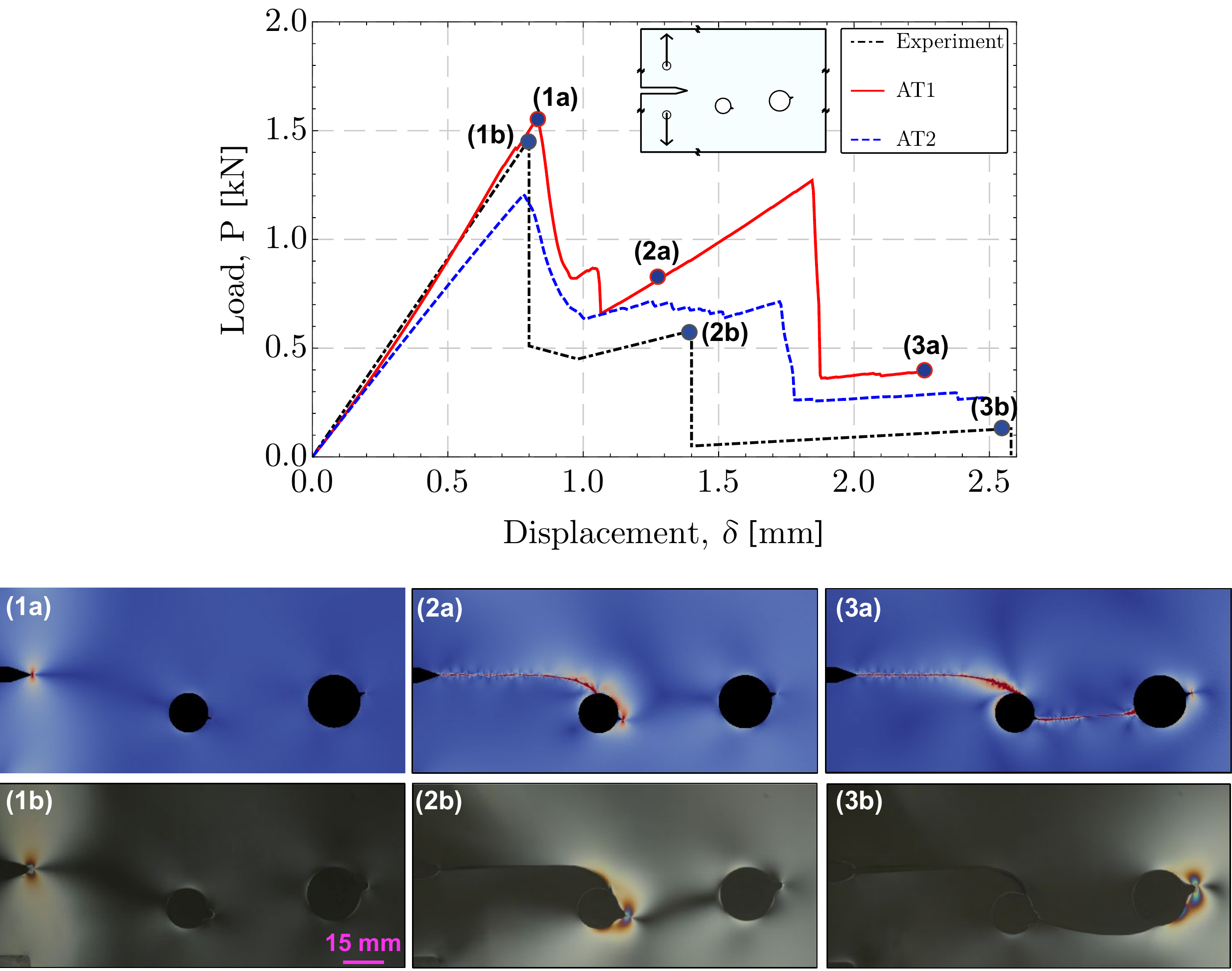}
 \caption{Comparison between simulation and experiment during the modified compact test 04 on a V-notched PMMA plate containing two holes. Force/displacement relation (upper part); crack path and photoelastic fringes at different displacements: simulation (central part) and experiment (lower part, from Fig. \ref{test_3_4b}). The numbers reported in the labels of the snapshots of the figure (central and lower parts) correspond to the respective points highlighted on the load/displacement graph.}
 \label{test_4_phase}
\end{figure}

The same numerical-experimental comparison proposed in the previous section is reported in Figs. \ref{test_3_phase} and \ref{test_4_phase}, for the
modified compact tests 03 and 04, containing now two circular holes with internal notches. The simulated crack trajectories according to the AT1 and AT2 models  reproduce very well the experiments. In the simulation of Test 03, the crack trajectory is attracted by the first hole and then starts propagating from the internal V-notch inclined at $30^{\circ}$, without reaching the second hole. In the simulation of Test 04, the crack hits the first hole and then the presence of a V-notch inclined at $15^{\circ}$ induces a secondary crack that is attracted by the second hole. Once the crack hits the second hole, it starts developing a tertiary crack from the notch of the second hole.

In terms of capability of reproducing the force-displacement curves, both AT1 and AT2 models satisfactorily predict the pre-peak branch, with AT1 slightly over performing the AT2. They also qualitatively predict in a correct way the shape of the post-peak response, while they present some difficulties in capturing the severe drops in the load carrying capacity observed in the experiments when crack propagates from the V-notch and, subsequently, from the first circular hole. Here the load level is overestimated.


\subsection{Phase field simulations of crack trajectories under bending}

Two numerical simulations are performed to simulate the four-point bending experiments shown in 
Figs. \ref{exp:4PBT1} and \ref{exp:4PBT2} on notched beams containing a circular hole. The difference between the two samples relies on the eccentricity of the circular hole with respect to the central notch.\\ 
The AT1 model accurately reproduces the mechanical response of the experiment, in terms of force-displacement curves, by setting $G_c=0.4$ N/mm, and $l=0.4$ mm. \\
In a similar fashion, the AT2 model provides a very good prediction of the force-displacement curves by setting $G_c=0.4$ N/mm
and $l=0.1$ mm for the first test, while $l=0.3$ mm for the second test. Such a slight tuning of the regularization parameter was necessary to simulate the sharp drop in the force-displacement curves as observed in the experiments. A video of simulations is available in the complementary material (movie SM4).

\begin{figure}[H]
 \centering
 \hspace{9mm}
 \includegraphics[width=1\linewidth]{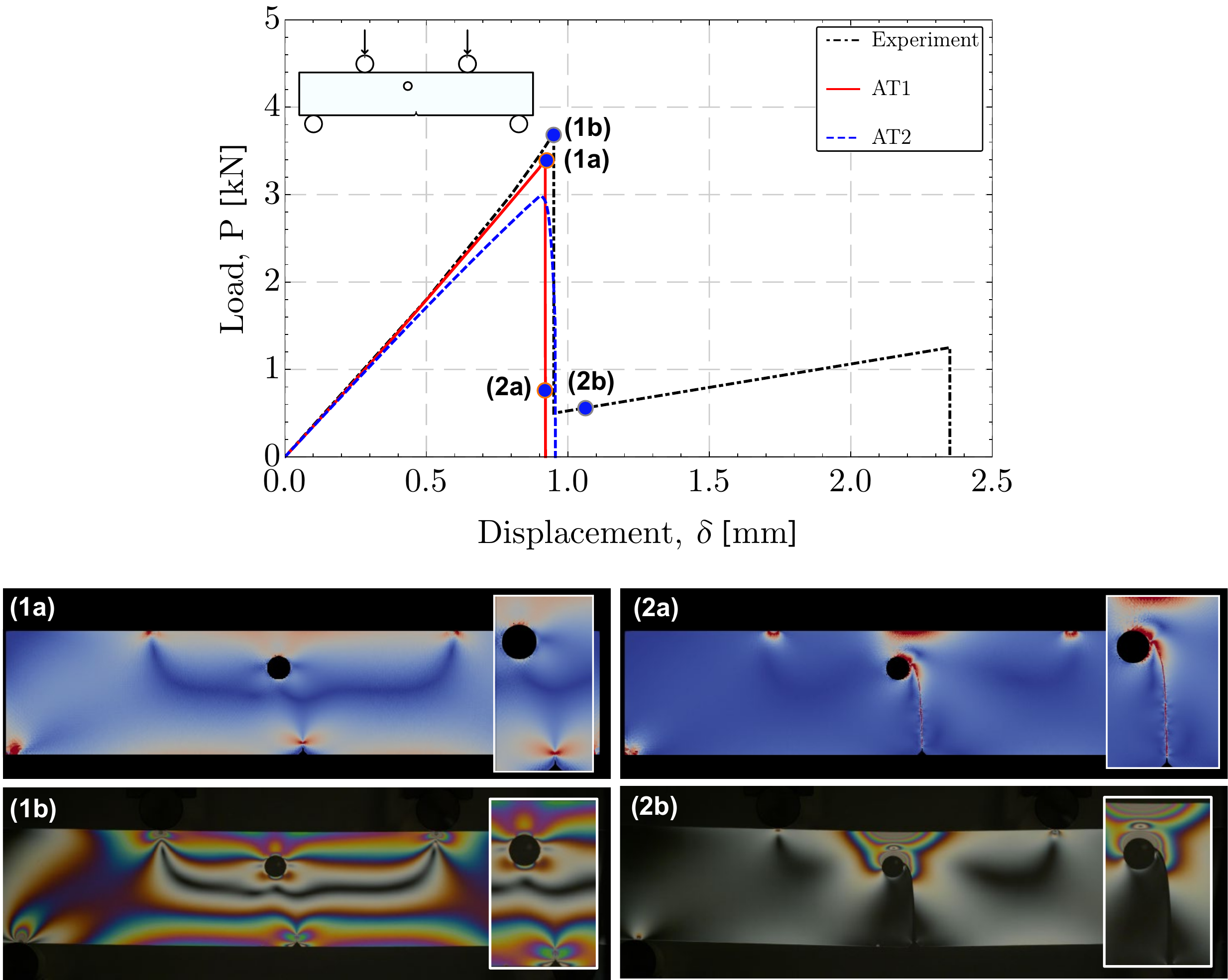}
 \caption{
 Comparison between simulation and experiment during the four-point bending test 01 on a notched beam containing a circular hole. Force/displacement relation (upper part); crack path and photoelastic fringes at two different displacements: simulation (central part) and experiment (lower part, from Fig. \ref{exp:4PBT1}). The numbers reported in the labels of the snapshots of the figure (central and lower parts)  correspond to the respective points highlighted on the load/displacement graph.
 }
 \label{simu:4PBT1_PF}
\end{figure}

\begin{figure}[H]
 \centering
 \hspace{9mm}
 \includegraphics[width=1\linewidth]{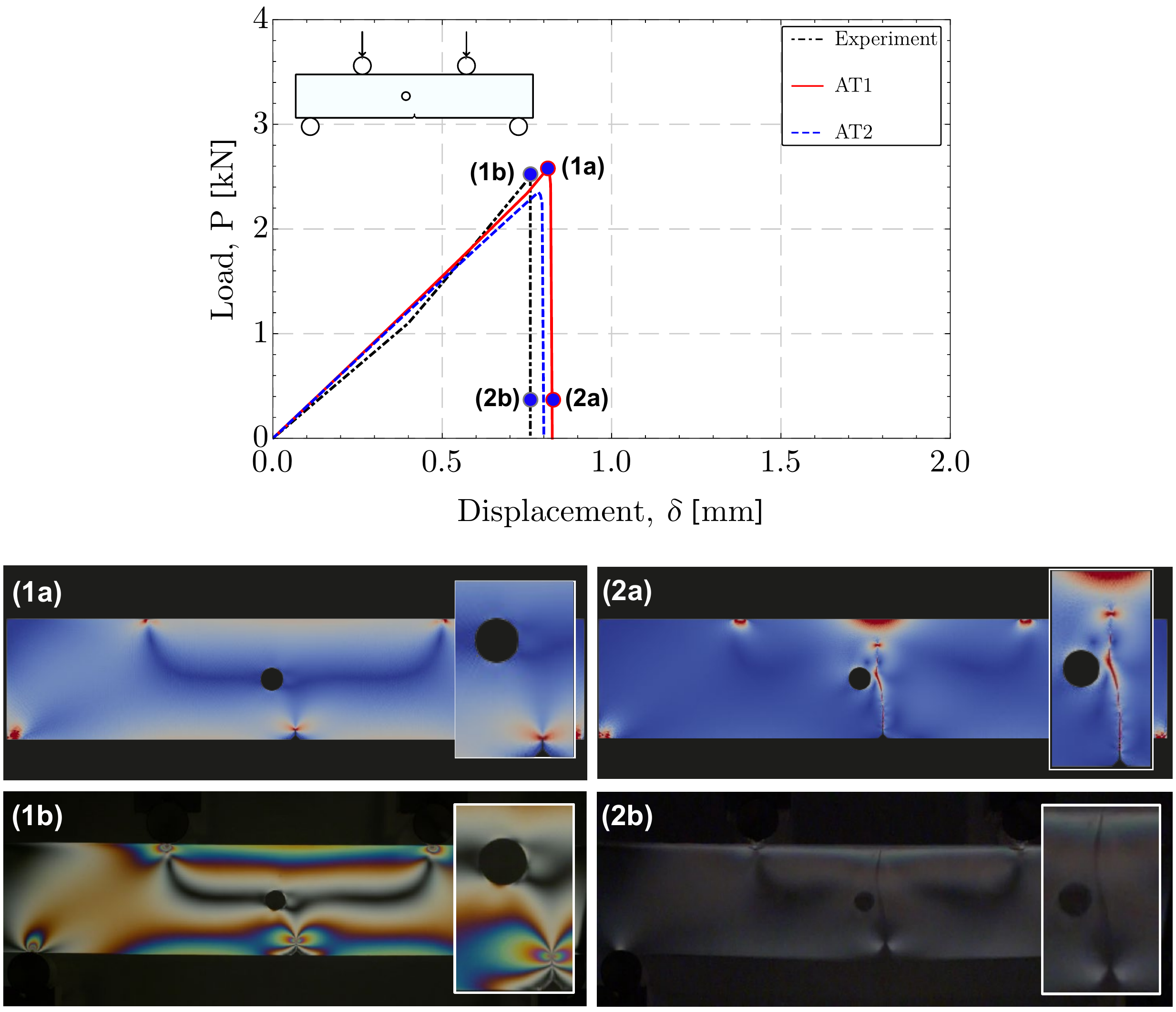}
 \caption{
 Comparison between simulation and experiment during the four-point bending test 02 on a notched beam containing a circular hole. Force/displacement relation (upper part); crack path and photoelastic fringes at two different displacements: simulation (central part) and experiment (lower part, from Fig. \ref{exp:4PBT2}). The numbers reported in the labels of the snapshots of the figure (central and lower parts) correspond to the respective points highlighted on the load/displacement graph.
}
 \label{simu:4PBT2_PF}
\end{figure}

In terms of crack paths, the phase field models are very accurate in predicting the trajectories and the major sequences of failure events. Fig. \ref{simu:4PBT1_PF} shows the simulated crack trajectory compared to the experimental one for the first test. Initially, the trajectory is straight, but later deviates towards the left due to the attraction by the circular hole, where it eventually stops. In the second test, Fig. \ref{simu:4PBT2_PF}, the crack trajectory is only deviated by the presence of the hole, but then it continues along vertical direction until it reaches the top side of the beam.


\subsection{Phase field simulations of crack trajectories under compression}

\begin{figure}[H]
 \centering
 \hspace{-10mm}
 \includegraphics[width=\linewidth]{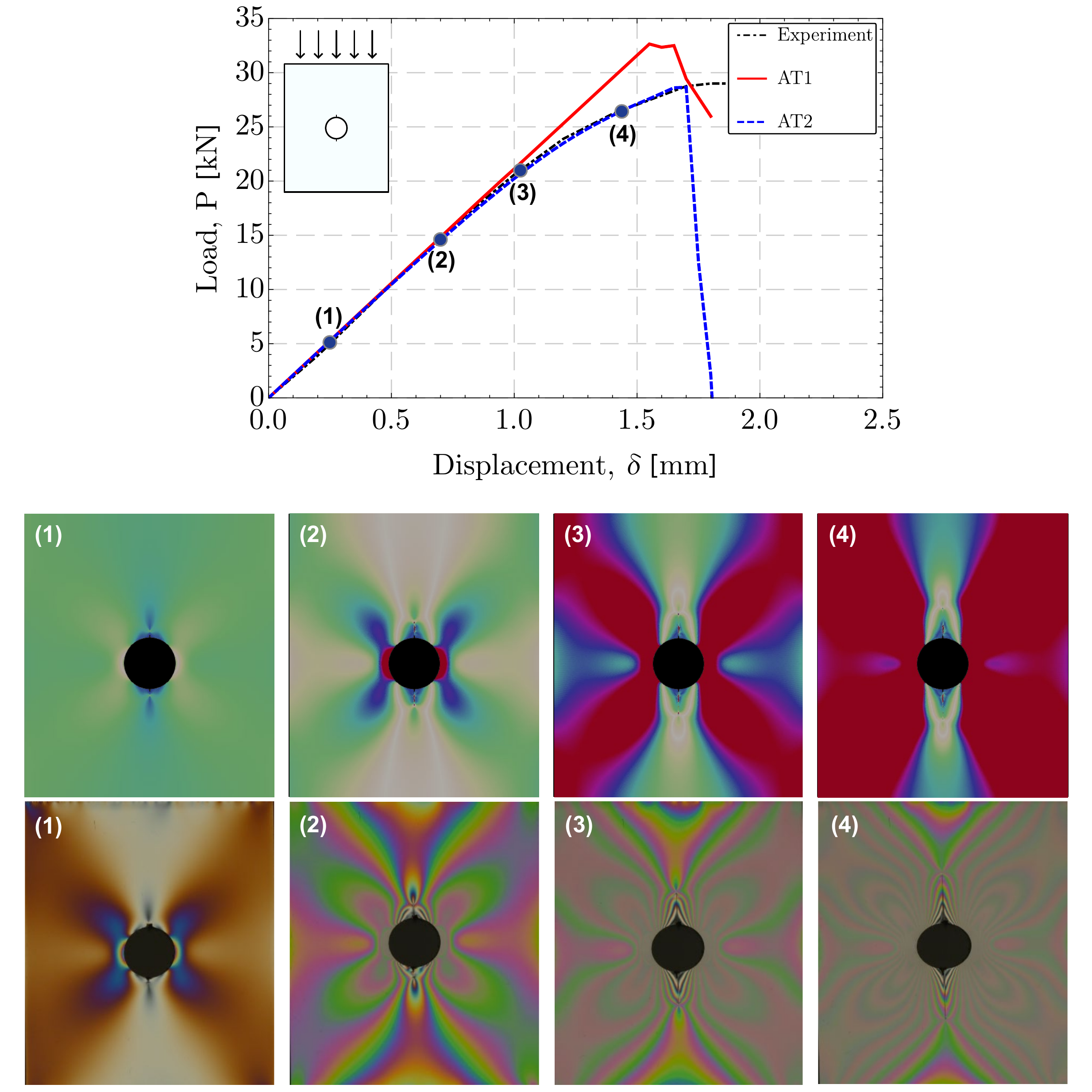}
 \caption{
 Comparison between simulation and experiment during a compression test of a PMMA plate containing a circular void with vertical pre-cracks. Force/displacement relation (upper part); crack path and photoelastic fringes at different displacements: simulation (central part) and experiment (lower part, from Fig. \ref{exp:compression}). textcolor{red}{The numbers reported in the labels of the snapshots of the figure (central and lower parts)  correspond to the respective points highlighted on the load/displacement graph.}
}
 \label{simu:COMP_FvsD}
\end{figure}

Two final numerical simulations are now reported of the experiments shown in Fig. \ref{exp:compression} on samples containing a circular hole under compression. A similar numerical simulation can be found in \cite{Tang}.
The comparison between numerically predicted and experimentally measured force-displacement curves shown in the upper part of Fig. \ref{simu:COMP_FvsD} highlights an excellent capability of both AT1 and AT2 models in providing quantitative predictions. To achieve that, a $G_c=0.4$ N/mm, plus a regularization parameter $l=0.4$ mm for the AT1 model and $l=0.4$ mm for the AT2 model.
In Fig. \ref{simu:COMP_FvsD} the crack path and the contour plots of the stress difference $\sigma_1 - \sigma_2$ resulting from the numerical simulation (central part of the figure) are compared with the experimentally observed photoelastic fringes (lower part of the figure), at the same selected points labeled on the force-displacement curve. Both phase field models are able to represent the spontaneous nucleation of cracks developing at
the vertical edges of the hole and evolving in the direction parallel to the imposed load. The load-displacement curve is also accurately reproduced by numerical simulations. A video of simulations is available in the complementary material (movie SM4).

\section{Conclusions}

Photoelastic experiments of brittle crack growth in specimens made of PMMA with different geometries, including (i) modified compact tests containing V-notches and one or two holes, (ii) plates with a hole and two sharp V-notches in compression, (iii) beams with a V-notch and a circular hole under four-point bending, have been proposed to specifically induce crack deflection and realize curved crack patterns caused by the rotation of the principal stress axes, occurring for stress states varying from pure tensile to tensile/compressive. 

The experimental results have been exploited to test 
the effectiveness of simulating  crack paths, fringe patterns, and even force-displacement curves, through the AT1 and AT2 phase field approaches to brittle fracture. 
As compared to the formulations examined in \cite{Tanne}, the models have been equipped with a  method based on the spectral decomposition of the strain tensor, originally proposed in  \cite{Miehe2010} to simulate fracture only in tension.

Overall, both AT1 and AT2 phase field models have been shown to accurately simulate the development of the complex crack trajectories observed in the experiments and the interaction between cracks and holes. In terms of ability to provide a quantitative prediction of the mechanical response, the AT1 and AT2 models have been shown to accurately simulate the measured  force-displacement curves in both the compressive tests on samples with a hole and the four-point bending tests. On the other hand, simulations evidenced some difficulties in capturing the sudden drop in the load carrying capacity occurring in the compact tests. It is relevant to remark that the latter tests (as simulated in the present article) significantly differ from those  addressed in \cite{Ambati2014}, where the specimen geometry and the type of loading were selected to be close to a set-up  corresponding to a uniaxial tensile test.\\
In this work, we attempted to identify the best parameters for the AT1 and AT2 models to match the experimental curves of the experimental tests done in conditions far from the
uniaxial stress state on which formulae in Eq. (17) are based. We found that the best parameters for the AT1 model are generally close to those expected
according to Eq. (17), which implies that the AT1 model has a good predictive capability just based on the estimate provided by Eq. (17). On the other hand, the AT2 model
required the use of $l$ much larger than the value predicted according to Eq. (17) to match the experimentally observed maximum loads. 

Finally, it is worth remarking that experiments have been designed to show that it is possible to \lq engineer' the propagation of a crack, so to obtain nucleation, propagation, arrest, with creation of secondary and tertiary trajectories. Therefore, the proposed results show that the developed experimental and numerical methodologies may lead to a new understanding of mechanical problems involving fracture growth and its possible control. 

\section*{Acknowledgements}
D.B. and R.C. gratefully acknowledge 
financial support from 
ERC-ADG-2021-101052956-BEYOND. M.P. and P.L. gratefully acknowledge funding from the Italian Ministry of University and Research (MIUR) to the project of national interest (PRIN 2017) XFAST-SIMS: Extra fast and accurate simulation of complex structural systems (grant agreement No. 20173C478N). D.M. gratefully acknowledges funding from H2020-MSCA-ITN-2020-LIGHTEN-956547.

\bibliographystyle{elsarticle-num}

\end{document}